\documentclass[aps,pra,twocolumn,showpacs,superscriptaddress]{revtex4-1}
\usepackage[english]{babel}
\usepackage{graphicx}
\usepackage{amsmath}
\usepackage{amssymb}
\usepackage{epstopdf}
\usepackage{amsfonts}
\usepackage{braket}
\usepackage{dcolumn}
\usepackage[breaklinks,colorlinks,linkcolor=blue,citecolor=blue,urlcolor=blue]{hyperref}
\usepackage[hyphenbreaks]{breakurl}

\DeclareMathOperator{\erfc}{erfc}
\usepackage{dcolumn}
\newcolumntype{d}[1]{D{.}{.}{#1}}
\begin{document}
\title{
  Incorporating exact two-body propagators for zero-range 
  interactions into $N$-body Monte Carlo simulations
}
\author{Yangqian Yan}
\affiliation{Department of Physics and Astronomy,
Washington State University,
  Pullman, Washington 99164-2814, USA}
\author{D. Blume}
\affiliation{Department of Physics and Astronomy,
Washington State University,
  Pullman, Washington 99164-2814, USA}
\date{\today}
\begin{abstract}
Ultracold atomic gases are, to a very good approximation, described by
pairwise zero-range interactions. This paper demonstrates that $N$-body
systems with two-body zero-range interactions can be treated reliably and
efficiently by the finite temperature and ground state path integral Monte
Carlo approaches, using the exact two-body propagator for zero-range
interactions in the pair product approximation. Harmonically trapped one- and
three-dimensional systems are considered. A new propagator for the
harmonically trapped two-body system with infinitely strong zero-range
interaction, which may also have applications in real time evolution schemes,
is presented.
\end{abstract}
\maketitle

\section{Introduction}
Systems with two-body zero-range interactions constitute important models in
physics. Although realistic two-body interactions do typically have a finite
range, results for systems with zero-range interactions provide a starting
point for understanding complicated few- and many-body dynamics. In
1934~\cite{fermi34},
Fermi used the zero-range model in quantum mechanical calculations to explain
the scattering of slow neutrons off bound hydrogen atoms.
Nowadays, the two-body contact interaction is discussed in elementary quantum
texts~\cite{claudequantum}. It has, e.g., been used to gain insights into the correlations of
molecules, such as $\text{H}_2^+$ and $\text{H}_2$, and to model atom-laser interactions~\cite{demkov88,janev91,fabrikant00}.

In the 60s~\cite{girardeau60,lieb63,gaudin67,yang67,yang69}, zero-range interactions were used extensively to model
strongly-interacting one-dimensional systems at zero and finite temperature.
Many of these models are relevant to electronic systems where the screening of
the long-range Coulomb interactions leads to effectively short-range
interactions~\cite{jones1985theoretical}. More recently, ultracold atomic gases interacting through
two-body van der Waals potentials have been, in the low temperature regime,
modeled successfully using zero-range
interactions~\cite{blochreview,stringariBosereview,stringariFermireview}. One-, two-, and
three-dimensional systems have been considered.

While zero-range interactions have been at the heart of a great number of
discoveries, including the Efimov effect~\cite{Efimov,fedorov94,BraatenHammerReview}, their incorporation into numerical
schemes is not always straightforward. Loosely speaking, the challenge in
using zero-range interactions in numerical schemes that work with continuous
spatial coordinates stems
from the fact that we, in general, do not 
know how to incorporate the boundary conditions implied
by the zero-range potential into numerical approaches at the four- and
higher-body level. 

This paper discusses an approach that allows for the use of
zero-range potentials in many-body simulations.
We work in position space and consider a system with fixed
number of particles.
We develop a scheme to incorporate pairwise zero-range interactions 
into $e^{-\tau H}$ directly, where $H$ is the system Hamiltonian.
The quantity $e^{-\tau H}$ is of fundamental importance. If $\tau$ is
identified with $1/(k_BT)$, where $k_B$ and $T$ denote the Boltzmann constant
and temperature, respectively, then $e^{-\tau H}$ is the density matrix for
the system at finite temperature. Knowing the density matrix, the
thermodynamic properties can be calculated. 
If, on the other hand, $\tau$ is identified with $it/\hbar$, where $t$ denotes
the real time, then  $e^{-\tau H}$ 
can be interpreted as the real time propagator and be
used to calculate dynamic properties.
Throughout this paper, we refer to $\tau$ as imaginary time, keeping in mind
that $\tau$ carries units of $1/\text{energy}$ and that $ \tau$ can be
associated with inverse temperature or real time.

The remainder of this paper is organized as follows. Section II reviews the
pair product approximation, which relates the many-body propagator to the
two-body propagator. Section III derives the
two-body propagator for various systems with zero-range interactions.
Sections IV and V demonstrate that the two-body zero-range propagators yield
reliable results if used in one- and three-dimensional path integral Monte
Carlo (PIMC)~\cite{ceperleyrev,boninsegni05,krauth2006statistical} and path integral ground state (PIGS)~\cite{ceperleyrev,pigs99,pigs00,massimo05,pigs08}
simulations of trapped
$N$-atom systems. The performance and implementation details will be
discussed.
While the free-space zero-range propagators have been reported in the
literature~\cite{schulman86,blinder88,lawande88,wodkiewicz91}, the zero-range propagators for the harmonically trapped
system with infinite coupling constant are, to the best of our knowledge, new.
Finally, Sec. VI concludes.

\section{$N$-body density matrix}
We consider $N$ particles with mass $m_j$ and position vector
$\mathbf{r}_j$ $(j=1,\dots,N)$ interacting via a sum of zero-range 
potentials with interaction strength $g$.
The Hamiltonian $H$ of the system can be written as
\begin{equation}
  H= \sum_{j=1}^N H_j^{\text{sp}} + \sum_{j<k}^{N} V_{jk},
  \label{Hamiltonian}
\end{equation}
where $H_j^{\text{sp}}$ denotes the non-interacting single-particle Hamiltonian of the $j$th particle
and $V_{jk}$ the two-body potential between the $j$th and $k$th particle.
In the following, it will be convenient  to separate the Hamiltonian $H_{jk}$,
where
$H_{jk}=H_j^{\text{sp}}+H_k^{\text{sp}}+V_{jk}$, of atoms $j$ and $k$ into relative and center of
mass pieces, $H_{jk}=H_{jk}^{\text{rel}}+H_{jk}^{\text{c.m.}}$,
where $H_{jk}^{\text{rel}}$ depends on the 
relative vector $\mathbf{r}_{jk}$, and 
$H_{jk}^{\text{c.m.}}$ on the center of mass vector 
$\mathbf{r}_{jk}^{\text{c.m.}}$, 
$\mathbf{r}_{jk}=\mathbf{r}_j-\mathbf{r}_k$
and $\mathbf{r}_{jk}^{\text{c.m.}}=(m_j\mathbf{r}_j+m_k\mathbf{r}_k)/(m_j+m_k)$.
Below, the non-interacting two-particle system will serve as a reference
system and we define $H_{jk}^0=H_j^{\text{sp}}+H_k^{\text{sp}}$ and
$H_{jk}^0=H_{jk}^{\text{rel},0}+H_{jk}^{\text{c.m.}}$.

The $N$-particle density matrix $\rho_{\text{tot}}(\mathbf{R},\mathbf{R}';\tau)$ in position space can be
written as
\begin{equation}
\rho_{\text{tot}}(\mathbf{R},\mathbf{R}';\tau)=
\Braket{\mathbf{R}|e^{-\tau H}|\mathbf{R}'},
  \label{}
\end{equation}
 where $\mathbf{R}=(\mathbf{r}_1,\dots,\mathbf{r}_N)$ and
 $\mathbf{R}'=(\mathbf{r}_1',\dots,\mathbf{r}_N')$ collectively denote the
 coordinates of the $N$-particle system.
 For sufficiently small $\tau$, $\rho_{\text{tot}}(\mathbf{R},\mathbf{R}';\tau)$
can be constructed using the
pair-product approximation~\cite{ceperleyrev}, 
\begin{eqnarray}
  \rho_{\text{tot}}(\mathbf{R},\mathbf{R}';\tau)
  \approx
  \left(\prod_{j=1}^N\rho^{\text{sp}}(\mathbf{r}_j,\mathbf{r}'_j;\tau)
    \right)\times\nonumber \\
  \left(\prod_{j<k}^N\bar{\rho}^{\text{rel}}(\mathbf{r}_{jk},\mathbf{r}'_{jk};\tau)
    \right)
    ,
  \label{rhotot}
\end{eqnarray}
where $\bar{\rho}^{\text{rel}}(\mathbf{r}_{jk},\mathbf{r}'_{jk};\tau)$
 denotes the normalized pair density matrix,
\begin{eqnarray}
  \bar{\rho}^{\text{rel}}(\mathbf{r}_{jk},\mathbf{r}'_{jk};\tau)
  =\frac{\rho^{\text{rel}}(\mathbf{r}_{jk},\mathbf{r}'_{jk};\tau)}
  {\rho^{\text{rel},0}(\mathbf{r}_{jk},\mathbf{r}'_{jk};\tau)},
  \label{}
\end{eqnarray}
and $\rho^{\text{rel}}(\mathbf{r}_{jk},\mathbf{r}'_{jk};\tau)$ and $\rho^{\text{rel},0}(\mathbf{r}_{jk},\mathbf{r}'_{jk};\tau)$ the relative density
matrices of the interacting and non-interacting two-body systems,
\begin{equation}
  \rho^{\text{rel}}(\mathbf{r}_{jk},\mathbf{r}'_{jk};\tau)=
\Braket{\mathbf{r}_{jk}|e^{-\tau H_{jk}^{\text{rel}}}|\mathbf{r}'_{jk}}
  \label{}
\end{equation}
and
\begin{equation}
  \rho^{\text{rel},0}(\mathbf{r}_{jk},\mathbf{r}'_{jk};\tau)=
  \Braket{\mathbf{r}_{jk}|e^{-\tau H_{jk}^{\text{rel},0}}|\mathbf{r}'_{jk}}.
  \label{}
\end{equation}
In Eq.~(\ref{rhotot}), $\rho^{\text{sp}}(\mathbf{r}_j,\mathbf{r}'_j;\tau)$ denotes the single-particle
density matrix,
\begin{equation}
  \rho^{\text{sp}}(\mathbf{r}_j,\mathbf{r}'_j;\tau)
  =\Braket{\mathbf{r}_j|e^{-\tau H_{j}^{\text{sp}}}|\mathbf{r}_{j}'}.
  \label{}
\end{equation}
The key idea behind Eq.~(\ref{rhotot}) is that the one- and two-body density
matrices can, often times, be calculated analytically. Indeed, 
the non-interacting propagator is known in the literature both for the
free-space and harmonically trapped systems~\cite{pathria1996statistical,ceperleyrev}.
Moreover, the eigen energies and eigen states of the Hamiltonian 
$H_{jk}^{\text{rel}}$ have, for a class of two-body interactions,
compact expressions,  which enables 
the analytical evaluation of the 
relative two-body density matrix
in certain cases (see Sec.~\ref{twobodyreltativedensitymatrixsection}).

It should be noted that the pair product approximation is only
valid in the small $\tau$ limit since it does not account for 
three- and higher-body correlations.
For the real time dynamics, this means that the time step is limited by the
importance of $N$-body ($N>2$) correlations.
If $\tau$ is identified with $1/(k_BT)$, the pair product approximation is 
limited to high temperature.
In this case, the pair product approximation is analogous to a virial
expansion that includes the second-order but not the third-order virial
coefficient~\cite{huang1987statistical}.

\section{two-body relative density matrix}
\label{twobodyreltativedensitymatrixsection}
In the following, we consider one- and three-dimensional systems, without and
with external harmonic confinement, and discuss the evaluation of the relative
density matrix for zero-range interactions.
For notational simplicity, we leave off
the subscripts $j$ and $k$ throughout this section, i.e., we denote the
relative distance vector by $\mathbf{r}$ for the three-dimensional system and
$x$ for the one-dimensional system, respectively, 
and the relative part of the two-body
Hamiltonian by $H^{\text{rel}}$. 

\subsection{One-dimensional system}
The complete set of bound and continuum
states of $H^{\text{rel}}$ is spanned by $\psi_n$ with eigen energies $E_n$
and $\psi_{k}$ with energies $\hbar^2k^2/(2\mu)$,
where $\mu$ denotes the reduced two-body mass and $k$ the relative scattering
wave vector.
If the $\psi_n$ and $\psi_k$ are normalized according to 
\begin{equation}
  \int\!\psi_n^*(x)\psi_{n'}(x)\,\mathrm{d}x=\delta_{nn'}
  \label{}
\end{equation}
and
\begin{equation}
  \int\!\psi_{k}^*(x)\psi_{k'}(x)\,\mathrm{d}x=\delta(k-k'),
  \label{normcondition}
\end{equation}
then the relative density matrix $\rho^{\text{rel}}(x,x';\tau)$
can be written as~\cite{pathria1996statistical}
\begin{eqnarray}
\rho^{\text{rel}}(x,x';\tau)=\sum_n
\psi_n^*(x)e^{-\tau E_n}
\psi_n(x')+ \nonumber\\
\int_{0}^{\infty}\!
  \psi_{k}^*(x)e^{-\tau \hbar^2 k^2/(2\mu)}
  \psi_{k}(x')
  \,\mathrm{d} k.
  \label{twobodydensitymatrix1d}
\end{eqnarray}

{\em{Free-space system:}}
The relative Hamiltonian for the free-space system with zero-range
interaction can be written as
\begin{equation}
  H^{\text{rel}}=-\frac{\hbar^2}{2\mu}\frac{\partial^2}{\partial x^2}+ g \delta(x),
  \label{1dfreehamiltonian}
\end{equation}
 where $g$ denotes the coupling constant of the $\delta$-function potential.
For positive $g$,
the Hamiltonian given in Eq.~(\ref{1dfreehamiltonian})
does not support a bound
state and
the corresponding energy spectrum is continuous.
The symmetric and anti-symmetric scattering states with energy
$\hbar^2k^2/(2\mu)$ read~\footnote{
  Equation~(\ref{normcondition}) is to be interpreted as follows:
  $\int\![\psi_{k}^s(x)]^*\psi_{k'}^s(x)\,\mathrm{d}x=\delta(k-k'),$
  $\int\![\psi_{k}^a(x)]^*\psi_{k'}^a(x)\,\mathrm{d}x=\delta(k-k'),$
  and
  $\int\![\psi_{k}^a(x)]^*\psi_{k'}^s(x)\,\mathrm{d}x=0$ (due to symmetry).
}
\begin{eqnarray}
  \psi_{k}^s(x)&=& \frac{1}{\sqrt{\pi}}\sin\left(k|x|+\delta(k)\right)
  \label{1dcontinuewavefunction}
\end{eqnarray}
and
\begin{eqnarray}
  \psi_{k}^a(x)&=& \frac{1}{\sqrt{\pi}}\sin(k x),
\end{eqnarray}
respectively; $\delta(k)=\arctan\left[\hbar^2k/(g\mu)\right]$ is the
phase shift.
For negative $g$, the Hamiltonian additionally supports a bound state
with symmetric wave function 
\begin{eqnarray}
  \psi_{0}^s(x)&=& \sqrt{\frac{\mu |g|}{\hbar^2}}e^{-\mu |g x|/\hbar^2}
  \label{}
\end{eqnarray}
and energy $-g^2\mu/(2\hbar^2)$.
Integrating over the symmetric and anti-symmetric scattering states,
 and adding, for negative $g$, the additional bound state,
 one finds the normalized relative density matrix $\bar{\rho}_{\text{1D,free}}^{\text{rel}}$~\cite{schulman86,blinder88,lawande88,wodkiewicz91},
\begin{align}
  \bar{\rho}_{\text{1D,free}}^{\text{rel}}(x,x';\tau) =1- 
  \exp \left(-\frac{\mu  \left(x x'+\left|x x'\right| \right)}{ \tau  \hbar ^2}\right)\times
\nonumber\\
  \sqrt{\frac{\pi\mu  \tau }{2}}\frac{g}{\hbar} \erfc(u)\exp(u^2),
  \label{finiteg1d}
\end{align}
where $u=\mu  \left(\left| x'\right| +\left| x\right| +g \tau   \right)/  \sqrt{2\mu  \tau \hbar^2}$ and $\erfc$ is the complementary error function.
We emphasize that Eq.~(\ref{finiteg1d}) holds for positive and negative $g$.
  The corresponding relative non-interacting density matrix  reads 
  \begin{equation}
    \rho_{\text{1D,free}}^{\text{rel},0}(x,x';\tau)=\left(\frac{\mu}{2\pi \tau \hbar^2}\right)^{1/2}
    \exp\left(-\frac{\mu(x-x')^2}{2\tau \hbar^2}\right).
    \label{1dfreerhononi}
  \end{equation}
  The free-space propagator given in Eq.~(\ref{finiteg1d}) was employed in a
  PIMC study of the harmonically trapped spin-polarized two-component Fermi
  gas with negative $g$~\cite{ceperley08}.

  For large $|g|$, $u$ approaches $\sqrt{\mu \tau/2}g/\hbar$ and, using $\lim_{u\to\infty} \sqrt{\pi}u \erfc(u)\exp(u^2)=1$,
  Eq.~(\ref{finiteg1d}) reduces to
  \begin{align}
    \bar{\rho}_{\text{1D,free}}^{\text{rel}}(x,x';\tau) = \begin{cases}
      1-\exp\left(-\frac{2\mu x x'}{\tau\hbar^2}\right)& \text{for $x x'>0$} \\
      0 & \text{for $x x'\le0$}.
    \end{cases}
    \label{1dfreerhonormal}
  \end{align}
  Equation~(\ref{1dfreerhonormal}) suggests that the relative coordinate does
  not change sign during the imaginary time
  evolution.
  Since the interaction strength is infinitely strong, the two
  particles fully reflect during any scattering process, i.e., the
  transmission coefficient is zero. 
  This means that the initial particle ordering remains unchanged during the
  time evolution.
  This is a direct consequence of the Bose-Fermi duality of one-dimensional
  systems~\cite{girardeau60,olshanii98,girardeau04}. Specifically, the phase shift 
  of the symmetric wave function given in
  Eq.~(\ref{1dcontinuewavefunction}) goes to zero when $|g|\to\infty$, implying
  that the symmetric wave functions coincide, except for an overall 
  $\text{sgn}(x)$
  factor, with the anti-symmetric scattering wave functions of non-interacting
  fermions.
  The implications of the Bose-Fermi duality for Monte Carlo simulations of
  $N$-body systems with infinite $g$ is discussed in Sec.~\ref{onedsection}.

{\em{Trapped system:}}
  For two particles in a harmonic trap, the system
  Hamiltonian reads
\begin{equation}
  H^{\text{rel}}=
  -\frac{\hbar^2}{2\mu}\frac{\partial^2}{\partial x^2}
  + g \delta(x)+ \frac{1}{2}\mu \omega^2 x^2,
  \label{SystemHamiltonian}
\end{equation}
where $\omega$ denotes the angular trapping frequency.
The energy spectrum of $H^{\text{rel}}$ is discrete and the
eigen energies and eigen functions are known analytically in compact 
form~\cite{busch}.
These solutions can be used to evaluate
Eq.~(\ref{twobodydensitymatrix1d}) numerically.
  The corresponding relative non-interacting density matrix reads
  \begin{eqnarray}
    \rho_{\text{1D,trap}}^{\text{rel},0}(x,x';\tau)=  
    \left[2\pi \sinh(\tau \hbar \omega )a_{\text{ho}}^2\right]^{-1/2}\times
    \nonumber \\ 
    \exp\left(-\frac{(x^2+x'^2)\cosh(\tau\hbar\omega)-2x x'}{2\sinh(\tau \hbar \omega) a_{\text{ho}}^2}
      \right),
    \label{1dtrapdensitymatrixnoni}
  \end{eqnarray}
  where $a_{\text{ho}}$ denotes the harmonic oscillator length,
  $a_{\text{ho}}=\sqrt{\hbar/(\mu\omega)}$.
For fixed $\tau$ and finite $g$, one can then
tabulate $\bar{\rho}_{\text{1d,trap}}^{\text{rel}}(x,x';\tau)$ for
discrete $x$ and $x'$ using 
Eq.~(\ref{twobodydensitymatrix1d})
and use a two-dimensional
interpolation during the $N$-body simulation.
The infinite sum in Eq.~(\ref{twobodydensitymatrix1d})
can be truncated by omitting terms with $n>n_{\text{max}}$, where
$n_{\text{max}}$ is chosen such that the Boltzmann factor fulfills the
inequality $e^{-\tau E_n}\ll e^{-\tau E_0}$. The value of $n_{\text{max}}$
depends on the time step: smaller $\tau$ require larger $n_{\text{max}}$.

For infinite $g$, we were able to derive a compact analytical expression for
$\bar{\rho}_{\text{1D,trap}}^{\text{rel}}(x,x';\tau)$.
As $g$ goes to infinity, the probability
distribution of each even state coincides with that of an odd state, i.e.,
the system is fermionized.
The complete set of even and odd eigen states for $g=\infty$ 
can be written as
\begin{eqnarray}
\psi_{n}^s(x)&=& \phi_n(|x|)
\label{1dtrapwavefunctions}
\end{eqnarray}
and
\begin{eqnarray}
\psi_{n}^a(x)&=&\phi_n(x),
\label{1dtrapwavefunctiona}
\end{eqnarray}
where $\phi_n(x)$ is the non-interacting harmonic oscillator wave function,
\begin{equation}
  \phi_n(x)=(\sqrt{\pi}2^nn!a_{\text{ho}})^{-1/2}e^{-x^2/(2a_{\text{ho}}^2)}H_n(x/a_{\text{ho}}),
  \label{noninteractingbasis}
\end{equation}
 $H_n(x)$ denotes the Hermite polynomial of order $n$, 
and $n$ takes the values $1, 3, 5, 7, \dots$. The corresponding
  energies are $E_{n}=(n+1/2)\hbar\omega$ for both the symmetric
  and anti-symmetric states, i.e.,
  each energy level is two-fold degenerate.
  Using Eqs.~(\ref{1dtrapwavefunctions}) and (\ref{1dtrapwavefunctiona}) in
  Eq.~(\ref{twobodydensitymatrix1d}) and evaluating the infinite sum
  analytically,
  we find
  \begin{align}
    \bar{\rho}_{\text{1D,trap}}^{\text{rel}}(x,x';\tau) = \begin{cases}
      1-\exp\left(-\frac{ 2x x'}{\sinh(\tau \hbar\omega)a_{\text{ho}}^2}\right)& \text{for $x x'>0$} \\
      0 & \text{for $x x'\le0$}.
    \end{cases}
    \label{1dtrapdensitymatrix}
  \end{align}
  For $\tau \hbar\omega\ll1$, i.e., when 
  the trap energy scale is much smaller than $1/\tau$,
  the trap propagator [Eq.~(\ref{1dtrapdensitymatrix})]
  equals the free-space propagator [Eq.~(\ref{1dfreerhonormal})].

  To test the one-dimensional propagators for infinite $g$, we consider the
  Hamiltonian given in Eq.~(\ref{SystemHamiltonian}) and prepare an initial
  state using a linearly discretized spatial grid. Our aim is to determine the
  ground state wave function and energy by imaginary time propagation. Two
  approaches are used. First, the initial state is propagated using the exact
trap propagators [see Eqs.~(\ref{1dtrapdensitymatrix}) and
  (\ref{1dtrapdensitymatrixnoni})]. In this case, the error originates solely
  from the discretization of the spatial degree of freedom; indeed, we find
  that the energy approaches the exact ground state energy quadratically with
  decreasing grid spacing $\delta x$. Second, the initial state is propagated
  using the free-space propagator [see Eqs.~(\ref{1dfreerhonormal}) and
    (\ref{1dfreerhononi})]. 
    We apply the Trotter formula~\cite{trotter1959product} and move half of the
    trap potential to the left and half to the right of the free-space
    Hamiltonian.
    This is known as the primitive approximation~\cite{ceperleyrev}, 
    which is expected to yield a quadratic time step error since the 
    trap potential does not commute with the free-space Hamiltonian.
    The error is found to scale
    quadratically with both the time step and the grid spacing.
    For 
    $\tau=(50\hbar\omega)^{-1}$, $\delta x=\sqrt{2}a_{\text{ho}}/40$,
    and $x_{\text{max}}=-x_{\text{min}}=4\sqrt{2}a_{\text{ho}}$, where
    $x_{\text{min}}\le x \le x_{\text{max}}$,
    we obtain energies that deviate by $2.4\times10^{-5}\hbar\omega$ and
    $2\times 10^{-12}\hbar\omega$ for the
    free-space propagator and the trap propagator, respectively,
    from the exact ground state energy of $3\hbar\omega/2$.

  \subsection{Three-dimensional system}
  Because the $s$-wave zero-range potential is spherically symmetric,
  the relative orbital angular momentum
  operator commutes with the relative Hamiltonian.
  Correspondingly, we label the bound states
  $\psi_{nlm}$ with eigen energies $E_{nl}$
  and the continuum states $\psi_{klm}$ with
  energies $\hbar^2k^2/(2\mu)$
 by the relative orbital angular momentum quantum number $l$ and the 
 projection quantum number $m$.
 If the $\psi_{nlm}$ and $\psi_{klm}$
 are normalized according to 
\begin{equation}
  \int\!\psi_{nlm}^*(\mathbf{r})\psi_{n'l'm'}(\mathbf{r})\,\mathrm{d}\mathbf{r}=\delta_{nn'}\delta_{ll'}\delta_{mm'}
  \label{}
\end{equation}
and
\begin{equation}
  \int\!\psi_{klm}^*(\mathbf{r})\psi_{k'l'm'}(\mathbf{r})\,\mathrm{d}\mathbf{r}=\delta(k-k')\delta_{ll'}\delta_{mm'},
  \label{}
\end{equation}
the relative density matrix $\rho^{\text{rel}}(\mathbf{r},\mathbf{r}';\tau)$ can be written as~\cite{pathria1996statistical}
\begin{eqnarray}
  \rho^{\text{rel}}(\mathbf{r},\mathbf{r}';\tau)=\sum_{nlm}
  \psi_{nlm}^*(\mathbf{r})e^{-\tau E_{nl}}
  \psi_{nlm}(\mathbf{r}')+ \nonumber\\
\sum_{lm}\int_{0}^{\infty}\!
  \psi_{klm}^*(\mathbf{r})e^{-\tau \hbar^2 k^2/(2\mu)}
  \psi_{klm}(\mathbf{r}')
  \,\mathrm{d} k.
  \label{twobodydensitymatrix}
\end{eqnarray}

{\em{Free-space system:}}
  The Hamiltonian of the three-dimensional system in free space
  reads
  \begin{equation}
    H^{\text{rel}}=-\frac{\hbar^2}{2\mu} \nabla_{\mathbf{r}}^2+\frac{2\pi\hbar^2
    a_s}{\mu}\delta^{(3)}(\mathbf{r})\frac{\partial}{\partial r}r,
    \label{freespace3dh}
  \end{equation}
   where  $a_s$ is the $s$-wave scattering length.
   The second term on the right hand side of 
   Eq.~(\ref{freespace3dh}) is the regularized two-body zero-range
   pseudopotential~\cite{zerorangepotential}.
  The $l>0$ continuum states read
  \begin{equation}
    \psi_{klm}(\mathbf{r})=i^l \sqrt{\frac{2}{\pi}}k j_l(k r)  Y_{lm}(\hat{\mathbf{r}}),
    \label{}
  \end{equation}
  where the 
  $j_l$ and $Y_{lm}$ denote spherical Bessel functions of the first kind
  and spherical harmonics, respectively.
  The continuum states for the $s$-wave channel read
  \begin{equation}
    \psi_{k00}(\mathbf{r})=\frac{1}{\sqrt{2}\pi r}\sin(k r+\delta_s(k)),
    \label{}
  \end{equation}
  where $\delta_s(k)=\arctan(-a_s k)$ is the $s$-wave phase shift.
  For positive $a_s$, there exists an $s$-wave bound state with eigen function
  $\psi_{000}(\mathbf{r})=1/\sqrt{2\pi a_s r^2} \exp(-r/a_s)$ and eigen energy
  $-\hbar^2/(2\mu a_s^2)$.
  As is evident from the above eigen states, only the $s$-wave states are
  affected by the interactions.
  Thus, we construct 
  the relative interacting density matrix $\rho_{\text{3D,free}}^{\text{rel}}$
  by writing the non-interacting relative density
  matrix $\rho_{\text{3D,free}}^{\text{rel},0}$ and subtracting from it the
  non-interacting $s$-wave contribution and adding to 
  it the $s$-wave contribution
  for finite $a_s$.

  For negative $a_s$, there exist only continuum states and
  the density matrix can
  be expressed as
  \begin{align}
    \rho_{\text{3D,free}}^{\text{rel}}(\mathbf{r},\mathbf{r}';\tau)=&\rho_{\text{3D,free}}^{\text{rel},0}(\mathbf{r},\mathbf{r}';\tau)+ \nonumber \\
  \int_0^\infty \!
  e^{-\frac{\tau \hbar^2 k^2}{2\mu}}
  \frac{1}{2\pi^2 r r'}[&\sin(k r + \delta_s(k))\sin(k r'+\delta_s(k))- \nonumber\\
    &\sin(k r)\sin(k
  r')]
  \,\mathrm{d} k,
    \label{3dfreeexpression}
  \end{align}
  where the non-interacting relative density matrix reads
  \begin{equation}
    \rho_{\text{3D,free}}^{\text{rel,0}}(\mathbf{r},\mathbf{r}';\tau)=
  (2\pi \hbar^2 \tau/\mu)^{-3/2}
  e^{-\mu(\mathbf{r}-\mathbf{r}')^2/(2\hbar^2 \tau)}.
    \label{3dfreespaceintegral}
  \end{equation}
  The integral in Eq.~(\ref{3dfreeexpression}) can be done
  analytically~\footnote{See entry 3.954 of I. S. Gradshteyn and I. M. Ryzhik,
{\em{Table of Integrals, Series, and Products}},
6th Ed., Academic Press.} and
  the normalized relative density matrix reads \cite{lawande88,wodkiewicz91}
  \begin{align}
    \bar{\rho}_{\text{3D,free}}^{\text{rel}}(\mathbf{r},\mathbf{r}';\tau)= 1+\frac{\hbar^2
    \tau}{\mu r r'}
    \exp\left(-\frac{\mu r r'(1+\cos\theta)}{\hbar^2 \tau}\right)
    \times \nonumber \\
    \left(1+\frac{\hbar}{a_s}\sqrt{\frac{\pi\tau}{2\mu}}\erfc(v)\exp(v^2)\right),
    \label{3dfiniteg}
  \end{align}
 where $\cos\theta= \mathbf{r}\cdot\mathbf{r}'/(r r')$ and
 $v=[r+r'-\tau \hbar^2/(\mu a_s)]/\sqrt{2\tau\hbar^2/\mu}$.
  Adding the bound state contribution to Eq.~(\ref{3dfreeexpression}) [see the 
    first term on the right hand side of Eq.~(\ref{twobodydensitymatrix})] 
  for positive $a_s$, one finds Eq.~(\ref{3dfiniteg}), i.e.,
  the same propagator as for negative $a_s$~\cite{wodkiewicz91}.

  For $|a_s|=\infty$, Eq.~(\ref{3dfiniteg})
  simplifies to
  \begin{equation}
    \bar{\rho}_{\text{3D,free}}^{\text{rel}}(\mathbf{r},\mathbf{r}';\tau)= 1+\frac{\hbar^2
    \tau}{\mu r r'}\exp\left(-\frac{\mu r r'(1+\cos\theta)}{\hbar^2 \tau}\right)
    .
    \label{3dfreedensitymatrix}
  \end{equation}
  This propagator was recently used in a proof-of-principle diffusion Monte
  Carlo study of the homogeneous two-component Fermi gas at unitarity with
  zero-range interactions~\cite{schmidt14}.

  {\em{Trapped system:}}
  The Hamiltonian for two particles in a spherically symmetric harmonic trap with 
  $s$-wave scattering
  length $a_s$ reads
  \begin{equation}
    H^{\text{rel}}=-\frac{\hbar^2}{2\mu} \nabla_{\mathbf{r}}^2+\frac{1}{2} \mu \omega^2
    \mathbf{r}^2+\frac{2\pi\hbar^2
    a_s}{\mu}\delta^{(3)}(\mathbf{r})\frac{\partial}{\partial r}r.
    \label{}
  \end{equation}
  The non-interacting relative density matrix reads
  \begin{eqnarray}
    \rho_{\text{3D,trap}}^{\text{rel},0}(\mathbf{r},\mathbf{r}';\tau)=
    a_{\text{ho}}^{-3}\left[2\pi \sinh(\tau\hbar\omega)\right]^{-3/2}\times\nonumber \\
    \exp\left({-\frac{(\mathbf{r}^2+\mathbf{r}'^2)\cosh(\tau\hbar\omega)
    -2\mathbf{r}\cdot\mathbf{r}'}{2\sinh(\tau\hbar\omega)a_{\text{ho}}^2}}\right).
    \label{}
  \end{eqnarray}
  Similar to the free-space case,
  the relative interacting density matrix is obtained from the non-interacting
  density matrix with the difference of the $s$-wave
  eigen states and energies of the interacting and non-interacting systems
  added.
  For finite $a_s$, we were not able to evaluate the infinite sum
  analytically. Because of the rotational invariance, the infinite sum depends
  only on $r$ and $r'$ (and not the direction of the vectors $\mathbf{r}$ and
  $\mathbf{r}'$), allowing for an efficient tabulation of the reduced relative
  density matrix.
  For infinitely large $a_s$, we find an analytical expression. In this case, 
  the bound state wave functions that are affected by the $\delta$-function
  interaction can be written as $\sqrt{2}\phi_n(r)/\sqrt{4\pi r^2}$, where the 
  $\phi_n(r)$ are 
  defined in Eq.~(\ref{noninteractingbasis}) 
  with $x$ replaced by $r$ and $n=0,2,4,\dots$
  The relative two-body density matrix reads
  \begin{eqnarray}
    \rho_{\text{3D,trap}}^{\text{rel}}(\mathbf{r},\mathbf{r}';\tau)=\rho_{\text{3D,trap}}^{\text{rel},0}(\mathbf{r},\mathbf{r}';\tau)+ \nonumber \\
    \sum_{n=0}^{\infty}
    e^{-\tau (n+\frac{1}{2})\hbar\omega}
  \frac{(-1)^n}{2\pi r r'}\phi_n^*(r)\phi_n(r'),
    \label{}
  \end{eqnarray}
  where the even $n$ terms in the sum over $n$ 
  come from the $s$-wave states that are
  affected by the interactions
  and the odd $n$ terms come from the $s$-wave states of 
  the non-interacting system.   Performing the infinite sum, we find for the normalized relative density matrix
  \begin{align}
    \bar{\rho}_{\text{3D,trap}}^{\text{rel}}&(\mathbf{r},\mathbf{r}';\tau)= 1+
    \nonumber\\
    &\frac{a_{\text{ho}}^2}{r r'}\sinh(\tau\hbar\omega)\exp\left(-\frac{r r'(1+\cos\theta)}{ a_{\text{ho}}^2\sinh(\tau\hbar\omega)}\right).
    \label{3dtrappeddensitymatrix}
  \end{align}
  Setting $\tau\hbar\omega=0$, Eq.~(\ref{3dtrappeddensitymatrix}) reduces to
  Eq.~(\ref{3dfreedensitymatrix}), i.e., to the corresponding free-space
  expression.

  Equations~(\ref{1dfreerhonormal}) and (\ref{3dfreedensitymatrix}) show that
  the 
  one- and three-dimensional free-space propagators for systems
  with infinitely large $\delta$-function strength are characterized by the
  length $\sqrt{\tau \hbar^2/\mu}$, which is proportional to the de Broglie
  wave length. The trap propagators for infinitely large coupling constant
  [see Eqs.~(\ref{1dtrapdensitymatrix}) and (\ref{3dtrappeddensitymatrix})],
  in contrast, are characterized additionally by the harmonic oscillator
  length.
  For finite interaction strength, the coupling constant defines a second
  length scale for the free-space system and a third length scale for the
  trapped system.

  \section{one dimensional tests}
  \label{onedsection}
  This section incorporates the trapped two-body propagator into
  PIMC calculations
  for one-dimensional $N$-particle systems with pairwise zero-range
  interactions. We find that the conventional PIMC sampling
  approaches~\cite{ceperleyrev,boninsegni05}
  yield an efficient and robust description of one-dimensional systems with
  two-body zero-range interactions.

  As a first example, we consider $N$ distinguishable 
  harmonically trapped particles with mass $m$ in one spatial dimension
  with pairwise zero-range interactions of infinite strength.
  The $N$-particle system with infinitely large interaction strength
  is unique in that the particle statistics becomes irrelevant for local
  observables. For example, the
  energy is the same for $N$ identical bosons, $N$ identical
  fermions and $N$ distinguishable particles at any temperature
  provided all particles interact via two-body zero-range interactions.
  We employ the zero-range trap propagator together with the
  single-particle trap propagator.
  Symbols in Fig.
\begin{figure}
\centering
\includegraphics[angle=0,width=0.4\textwidth]{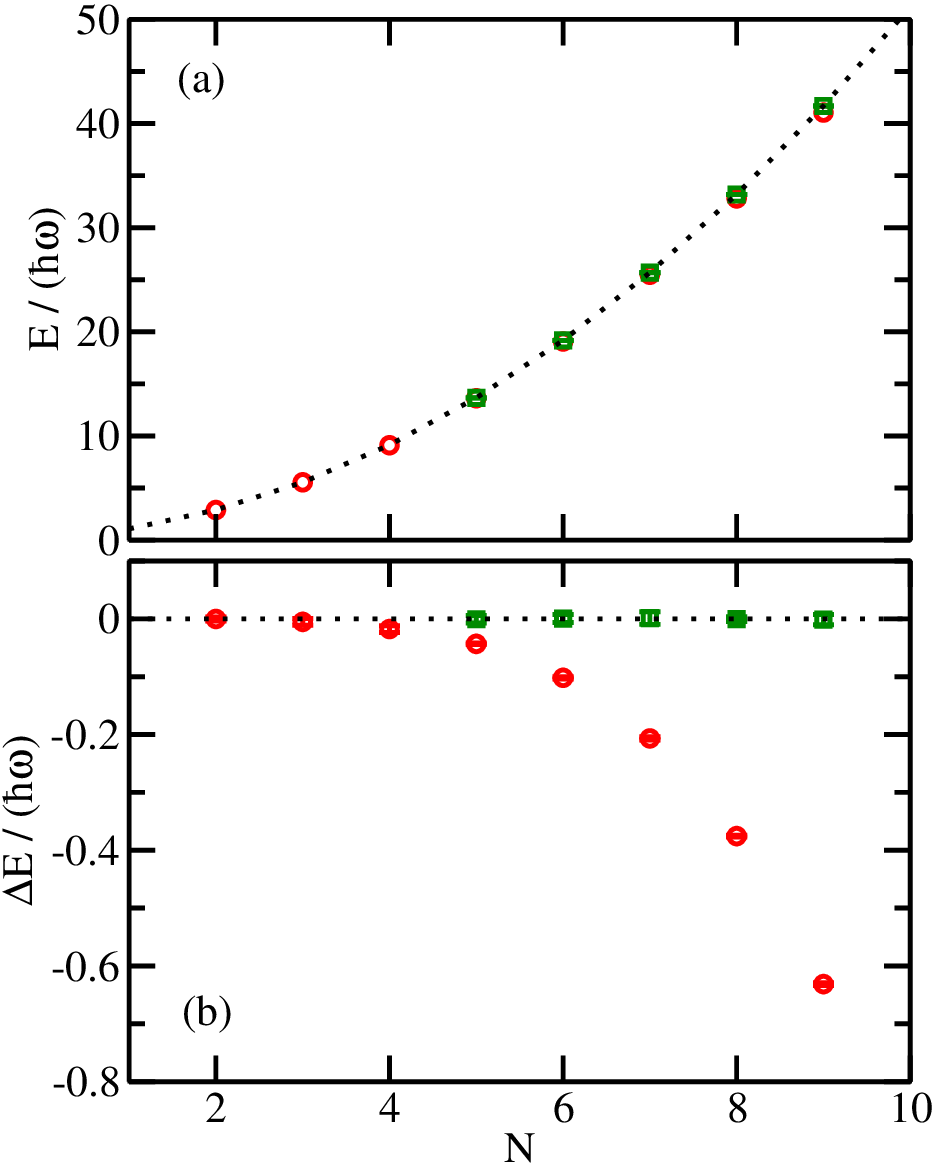}
\caption{(Color online)
  PIMC results for $N$ harmonically trapped distinguishable one-dimensional 
  particles with two-body
  zero-range interactions of infinite strength at temperature
  $T=\hbar\omega/k_B$.
  (a) The symbols show the energy obtained by the PIMC approach as a function
  of the number of particles $N$. For comparison, the dotted line shows the
  exact thermally averaged energy.
  (b) Symbols show the energy difference $\Delta E$ between the PIMC energies
  and the exact results. As a reference, the dotted line shows the $\Delta
  E=0$ curve.
  In (a) and (b), the circles and squares are calculated using $8$ and $128$
  time slices, respectively.
 }\label{Fig1}
\end{figure} 
  \ref{Fig1}(a) show the PIMC energy for $N$ distinguishable
  particles at temperature
  $k_BT=\hbar\omega$.
  Circles and squares are for simulations with imaginary time
  step  $\tau\hbar\omega=1/8$ and $1/128$, respectively.
  For comparison, the 
  dotted line is calculated directly from the partition function
  of $N$ non-interacting fermions.
  Figure \ref{Fig1}(b) shows the energy difference $\Delta E$ between 
  the simulation and the
  analytical results.
  It can be seen that the calculations for the larger time step (circles in
  Fig.~\ref{Fig1}) exhibit a systematic time step error, which
  is found to scale quadratically with the time step $\tau$ for fixed $N$ and
  to
  originate from the pair product approximation.
  Since we include the two-body correlations exactly, the leading order error
  is expected to come from three-body correlations.
  Indeed, for relatively small fixed $\tau$ ($\tau^{-1}$ around $16\hbar\omega$)
  and varying $N$, we find that the error $\Delta E$ scales approximately
  linearly with the number of
  triples, suggesting that the error is dominated by three-body correlations
  with sub-leading contributions arising from four-body correlations. 
  As the number of particles $N$ or the time step $\tau$ increase,
  four- and higher-body correlations
  become more important.
  This error analysis suggests that an improved propagator could be obtained
  if the three-body problem could be solved
  analytically.
  We note that the performance of the zero-range trap propagator for the
  system with infinite $g$ is quite similar to that of the free-space
  propagator using the second- or fourth-order Trotter decomposition. The
  reason is that the error is dominated by three- and higher-body
  correlations.

  We now discuss that the simulations need to be modified to treat $N$
  identical bosons or fermions with pairwise zero-range interactions of
  infinite strength.
  Equations~(\ref{1dfreerhonormal}) and (\ref{1dtrapdensitymatrix}) 
  indicate that the paths 
  for any two particles cannot cross. This implies that the 
  permute move, implemented following the approach
discussed in Ref.~\cite{massimo05}, yields a zero acceptance probability.
  This is consistent with our finite $g$ simulations for $N$ identical bosons.
As we change $g$ for otherwise fixed simulation parameters from small positive to large positive values,
  the probability of sampling the identity
  permutation approaches 1.
  The fact that particle permutations are always rejected,
causes two problems for the infinite $g$ simulations.
  First, since the particle ordering does not change, the 
single particle density for the first particle differs
from that of the second particle, and so on. To calculate the 
single particle density $\rho(x)$
of, e.g., the $N$ identical boson system,
one can average the single particle density $\rho_j(x)$ of the $j$th particle 
over all $j$, $\rho(x)=N^{-1} \sum_{j=1}^N \rho_j(x)$.
An analogous average can be performed for other local (closed paths) structural properties.
  Second, the simulation of open paths, which allow
for the calculation of off-diagonal long-range order,
requires that the sampling scheme be modified since 
  open paths do allow for permutations. The two-particle density matrix
  $\rho(\{x_1,x_2\},\{x_2,x_{1}'\};\tau)$, e.g., is finite if $x_1 < x_2 < x_1'$.
The calculation of non-local observables is beyond the scope of the present paper.

  As a next example, we apply the zero-range trap propagator to $N=2$ and $3$ 
  identical bosons in a harmonic trap with 
  $g=\hbar^2 /(\sqrt{2}\mu a_{\text{ho}})$. 
  For the PIMC calculations, we tabulate the density matrix for the time step of interest and
  interpolate during the simulation. For small number of particles, we expect 
  the zero-range trap propagator to work well even
  for a large time step and we use $\tau\hbar\omega=1/2$.
  For $k_BT=\hbar\omega/32$, we obtain an energy of 
  $1.3067(1)\hbar\omega$ and $2.3880(1)\hbar\omega$ for $N=2$ and $3$,
  respectively.
  The temperature is so low that the system is essentially in the
  ground state.
  For comparison, we determined the ground state energy using the
  transcendental equation from Ref.~\cite{busch} and by solving the 
  Lippmann-Schwinger equation~\cite{ebrahim12}.
  The resulting ground state energies [$1.306746\hbar\omega$ 
  and $2.3880(1)\hbar\omega$ for $N=2$ and $3$, respectively] 
  agree within error bars with the PIMC results.

  To demonstrate that the PIMC simulations describe the short-distance behavior
  of systems with zero-range interactions correctly, we analyze the 
  pair distribution function $P_{12}(x)$, which is normalized to 
  $\int_{-\infty}^{\infty}\!P_{12}(x)\,\mathrm{d}x=1$, 
  for $N=2$ and $3$ identical bosons with
  finite $g$.
  To start with, we derive the 
  short-distance properties of the pair distribution
function $P_{12}$ for $N$ identical bosons with zero-range interactions.
  Using the Hellmann-Feynman theorem,
  the partial derivative of the energy with respect to $g$ can be related to the
  pair distribution function at $x=0$~\cite{1dtan11},
  \begin{equation}
    P_{12}(0)=\frac{2}{N(N-1)} \frac{\partial E}{\partial g}.
    \label{1dcontact}
  \end{equation}
  Note that Eq.~(\ref{1dcontact}) is the one-dimensional analog of 
  equating the three-dimensional adiabatic and pair
  relations~\cite{tan1}.
  Second, from the Bethe-Peierls boundary condition of the $N$-body 
  wave function $\Psi$ (the derivatives are taken while all other coordinates
  are kept fixed),
  \begin{equation}
    \frac{\partial\Psi}{\partial x_{jk}}\bigg|_{x_{jk}\to0^+}-\frac{\partial \Psi}{\partial x_{jk}}\bigg|_{x_{jk}\to0^-}
    =\frac{2\mu g}{\hbar^2} \Psi \bigg|_{x_{jk}\to 0},
    \label{}
  \end{equation}
  one can derive that the slope of the pair distribution function 
  at any temperature
  satisfies
  \begin{equation}
    \frac{\partial}{\partial x}P_{12}(x)\big|_{x\to0^+}-
    \frac{\partial}{\partial x}P_{12}(x)\big|_{x\to0^-}
    =\frac{4\mu g}{\hbar^2}P_{12}(x)\big|_{x\to0}.
    \label{contact}
  \end{equation}
  For identical bosons, $\partial P_{12}(x)/\partial x\big|_{x\to0^+}$ and
    $\partial P_{12}(x)/\partial x\big|_{x\to0^-}$ have the same magnitude but opposite signs.

\begin{figure}
\centering
\includegraphics[angle=0,width=0.4\textwidth]{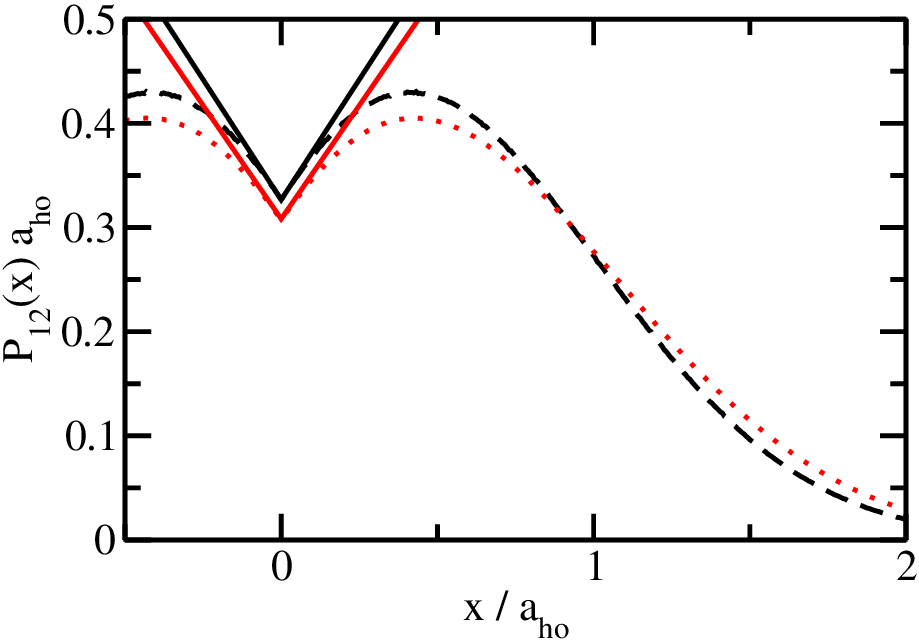}
\caption{(Color online)
  PIMC results for harmonically trapped one-dimensional bosons interacting
  through two-body
  zero-range interactions with coupling constant $g=\hbar^2/(\sqrt{2}\mu a_{\text{ho}})$
  at temperature $T=\hbar\omega/(32k_B)$.
  The dashed and dotted lines show the pair distribution function obtained by
  the PIMC approach for $N=2$ and $3$ bosons, respectively.
  For comparison, the solid lines show the asymptotic short-range
  behavior obtained by alternative means (see text).
 }\label{Fig2}
\end{figure}

  The dashed and dotted lines in Fig.~\ref{Fig2}
show $P_{12}(x)$ obtained from our PIMC simulation for $N=2$ and $3$,
respectively.
For comparison, the 
solid lines are obtained using Eqs.~(\ref{1dcontact}) and (\ref{contact}).
The values of $\partial E/\partial g$ are obtained through independent energy
calculations using the techniques of Refs.~\cite{busch,ebrahim12}.
  We find $P_{12}(0)=0.3266002/a_{\text{ho}}$ and $0.308245(2)/a_{\text{ho}}$
  for $N=2$ and $3$, respectively.
  Our PIMC results agree well with the solid lines in the small $|x|$ regime,
  demonstrating that the PIMC approach describes the short-range behavior
  accurately.

  \section{three dimensional tests}
  The pair distribution function of the non-interacting three-dimensional
  system is, unlike that of the non-interacting one-dimensional 
  system, zero at vanishing interparticle distance.
  This fact leads, as we discuss now, to non-ergodic behavior unless the
  traditional path integral sampling methods are complemented by
  an additional move. To motivate the introduction of this new ``pair distance
  move'', we consider the two-particle system.
\begin{figure}
\centering
\includegraphics[angle=0,width=0.4\textwidth]{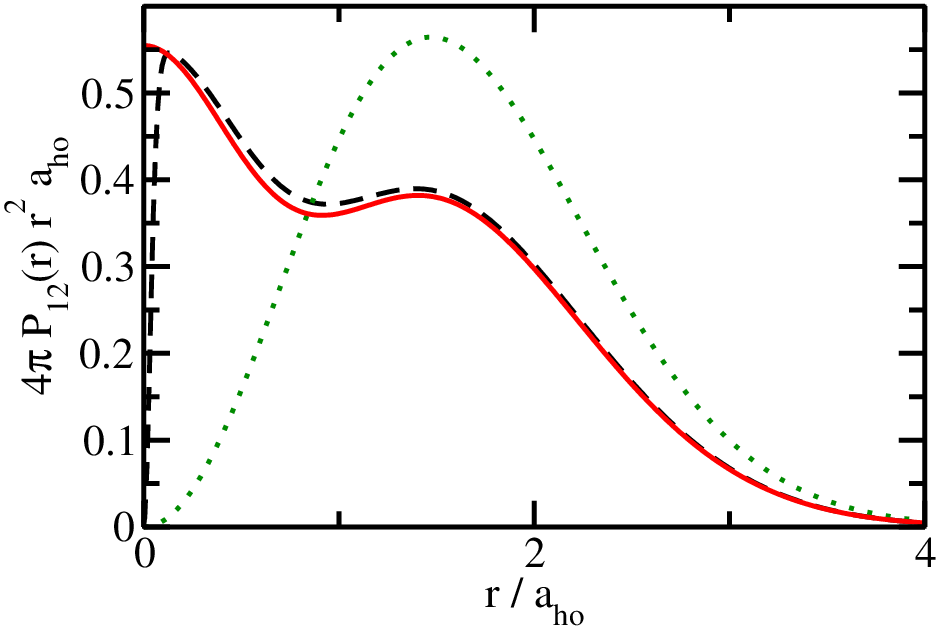}
\caption{(Color online)
  Scaled pair distribution functions $4\pi P_{12}(r) r^2$ 
  for two distinguishable particles of mass $m$
  in a harmonic trap at $k_BT/(\hbar\omega)=1$.
  The solid and dashed lines are for two particles with infinitely large
  $s$-wave scattering length interacting through the zero-range potential and
  a Gaussian potential with effective range $r_e\approx0.0861a_{\text{ho}}$,
  respectively.
  For comparison, the dotted line 
  is for the non-interacting system.
 }\label{Fig3}
\end{figure} 

  Solid and dashed lines in Fig.~\ref{Fig3} show  the scaled pair distribution functions
  for two distinguishable particles with infinitely large $s$-wave scattering
  length in a three-dimensional harmonic trap at
  $k_BT/(\hbar\omega)=1$
  interacting through a zero-range potential and a finite-range
  Gaussian potential with effective range $r_e\approx0.0861a_{\text{ho}}$,
  respectively.
  The pair distribution
  function $P_{12}(r)$ is normalized according to 
  $4\pi \int_0^{\infty}\! P_{12}(r)r^2\,\mathrm{d}r=1$.
  The pair distribution functions for 
  the finite-range and zero-range potentials nearly coincide for large $r$,
  but differ for small $r$.
  The pair distribution function for the Gaussian potential drops to 0 as $r$
  approaches 0 while that for the zero-range potential approaches a non-zero
  value.

  PIMC and PIGS simulations typically use the first term 
  on the right hand side of Eq.~(\ref{rhotot}) as the ``prior distribution''
  and the second term as the ``correction''.
  This is suitable for $N$-body systems with 
  two-body finite-range potentials
  for which the scaled pair distribution function is, as that of the
  non-interacting system (see the dotted line in Fig.~\ref{Fig3} for a
  two-body example),
  zero at $r=0$.
  Since the prior distribution has zero probability at $r=0$, the pair
  distribution function of the system with 
  zero-range interactions is not properly
  sampled if standard sampling schemes are used.
  Ergodicity is violated at $r=0$ and the probability to sample the region
  near $r=0$ is small.
  Moreover, the correction term [see Eqs.~(\ref{3dtrappeddensitymatrix}) and
  (\ref{3dfreedensitymatrix})]
  diverges as $r$ or $r'$ go to 0. This means that the probability to sample
  configurations with $r\approx 0$ is small.
  However, if such a configuration is
  chosen, there is a small probability to accept a new configuration with much
  larger $r$, i.e., the correlation length is large.
  Similarly, if one uses the naive uniform distribution for the prior
  distribution, i.e., if one proposes a move for which
  all Cartesian coordinates differ by $\delta x$ from the current
  configuration, where $\delta x$ is a random value between $-\Delta x$ and
  $\Delta x$, the problems discussed above still exist.

  To remedy the problems that arise if the standard sampling approaches are
  used, we introduce a ``pair distance move'' for which the prior distribution 
  scales as $1/r^2$ in the pair distances.
  First, particles $j$ and $k$ and a bead $l$ are chosen (the coordinates for
  the $l$th bead are collectively denoted by $\mathbf{R}_l$)
  and the distance $r_{jk}=|\mathbf{r}_{jk}|$ and the direction
  $\hat{\mathbf{r}}_{jk}$ are  calculated.
  A new vector $\mathbf{r}_{jk,\text{new}}$ that lies along
  $\hat{\mathbf{r}}_{jk}$ or $-\hat{\mathbf{r}}_{jk}$ is proposed, 
  $\mathbf{r}_{jk,\text{new}}=\epsilon \hat{\mathbf{r}}_{jk}$. The quantity $\epsilon$
  is written as $\epsilon=r_{jk}+\delta r$, where 
  $\delta r$ is obtained by choosing uniformly from $-\Delta r$ to $\Delta r$.
  If the weight $w$,
  \begin{equation}
    w=\min[1,\frac{\rho_{\text{tot}}(\mathbf{R}_{l-1},\mathbf{R}_{l,\text{new}};\tau)\rho_{\text{tot}}(\mathbf{R}_{l+1},\mathbf{R}_{l,\text{new}};\tau) \epsilon^2}
      {\rho_{\text{tot}}(\mathbf{R}_{l-1},\mathbf{R}_{l};\tau)\rho_{\text{tot}}(\mathbf{R}_{l+1},\mathbf{R}_{l};\tau) r_{jk}^2}],
    \label{}
  \end{equation}
  is larger than a uniform random number between 0 and 1, then the
  proposed move is accepted.
  Otherwise, it is rejected and the old configuration is kept. The value of
  $\Delta r$ is adjusted such that about $50\%$ of the proposed moves are
  accepted.
  It can be easily proven that detailed balance is fulfilled.
  Our PIMC calculations for the two-body system 
  with zero-range interactions show that the ``pair distance move'' significantly
  improves the sampling.
  Without this move, the short-range behavior of the pair
  distribution function has a long correlation length, which increases with
  decreasing $\tau$.
  With this move, the small $r$ behavior is described accurately.
  The move described here 
  is related to the compression-dilation move introduced 
  in Ref.~\cite{krauth13}.
  Few details were given in Ref.~\cite{krauth13} and no comparison with that
  approach is made in this paper.

We now demonstrate that the outlined sampling scheme provides a reliable
description of Bose systems at unitarity, which have attracted a great deal of
attention recently experimentally and theoretically~\cite{salomon13,hadzibabic13,jin14,krauth13,bohn14,zhou14,platter14,toigo14}. While the properties of
unitary Fermi systems with zero-range interactions are fully determined by the
$s$-wave scattering length~\cite{stringariFermireview,blochreview,blumerev12,schmidt03,giorgini04,castin06pra} those of Bose systems additionally depend on a
three-body parameter~\cite{Efimov,BraatenHammerReview}. Specifically, if the two-body interactions are modeled
by zero-range potentials, then a three-body regulator is needed to prevent the
Thomas collapse of the $N$-boson ($N\ge3$) 
system~\cite{thomas35,BraatenHammerReview}. Here, we utilize a purely
repulsive three-body potential of the form \begin{equation}
  V_{\text{3b}}(R_{jkl})=\frac{C_6}{R_{jkl}^6},
  \label{}
\end{equation}
where $R_{jkl}$ denotes the three-body hyperradius, $R_{jkl}=\sqrt{(r_{jk}+r_{jl}+r_{kl})/3}$.
In the $N$-boson system, each of the $N(N-1)(N-2)/6$ triples feels the
regulator, i.e., the term $\sum_{j<k<l}V_{\text{3b}}(R_{jkl})$ is added to the
Hamiltonian with pairwise zero-range interactions. In the absence of an
external trap, the zero temperature three-body ground state energy
$E_{\text{trimer}}$ of the unitary system is set by the $C_6$ coefficient. The
corresponding length scale is $1/\kappa$, where
$\kappa=\sqrt{m|E_{\text{trimer}}|}/\hbar$ is the binding momentum.

Our goal is to determine the ground state properties of self-bound $N$-boson
droplets at unitarity in the absence of an external confinement. In the context
of the present paper, it would seem that our goal could be readily achieved
using the PIGS approach. It turns out, however, that without a good initial
trial wave function, the number of time slices
needed to converge the calculations is rather large, making the simulations
computationally quite expensive. Instead, one might consider
performing PIMC calculations at various temperatures and extrapolating to the
zero temperature limit.
This approach also turns out to be computationally extensive. Our simulations
pursue an alternative approach, in which the scattering states of the system
are discretized in such a way that the relative ground state energy
$E_{\text{cluster}}$ of the $N$-body cluster is much larger than the energy
scale introduced by the discretization. We utilize a spherically symmetric
harmonic trap and adjust the trapping frequency such that
$|E_{\text{cluster}}|\gg\hbar\omega$. Simulations are then performed at a
temperature where the Bose droplet is in the ground-state dominated
liquid-phase~\cite{krauth13,gaussian14}, where the finite temperature introduces center-of-mass
excitations but not excitations of the relative degrees of freedom. The
temperature $T_{\text{tr}}$ 
at which excitations of the relative degrees of freedom become
relevant can be estimated using the ``combined model'' introduced in
Ref.~\cite{gaussian14}.
As we show now, this approach allows for a fairly
robust determination of the $N$-boson properties at zero temperature.

We set the trap energy $\hbar \omega$ to $0.27 |E_{\text{trimer}}|$
($E_{\text{trimer}}$ is the ground state energy of the three-boson system
in free space) and the temperature to $\hbar \omega / k_B$.
These parameters provide a good compromise: First, the temperature is 
sufficiently low that finite temperature effects are negligible
(i.e., $T < T_{\text{tr}}$ for the $N$ considered below, $N=3-9$)
and high enough that convergence can be reached with the computational
resources available to us. Second, the size of the $N$-boson system
is smaller than the harmonic oscillator length such that structural
properties such as the pair distribution function are largely 
unaffected by the external confinement for $N \gtrsim 5$.

Our path integral simulations use the two-body zero-range
trap propagator. The repulsive three-body potentials are treated
using the Trotter formula. In the second-order
scheme, half of the sum of the three-body potentials
is moved to the left and half to the right of the Hamiltonian $H$ that
accounts for the two-body interactions and the external confinement.
In the fourth-order scheme, a more involved decomposition is
used~\cite{chin97,voth}.
In addition to the standard moves and the ``pair distance move'', we 
implement a move that updates the center-of-mass coordinates.
The introduction of this center-of-mass
move is motivated by the fact that the
relative degrees of freedom are expected to be, to a 
good approximation, ``frozen'' in the ground state while 
low-energy center-of-mass excitations are allowed. Indeed, the
center-of-mass energy is given by $E_{\text{c.m.}}=
3\hbar\omega \coth(\hbar\omega/(2k_BT))/2$,
which evaluates to $3.24593 \hbar \omega$ for $T=\hbar \omega / k_B$,
indicating that center-of-mass excitations cannot be neglected.

\begin{figure}
\centering
\includegraphics[angle=0,width=0.4\textwidth]{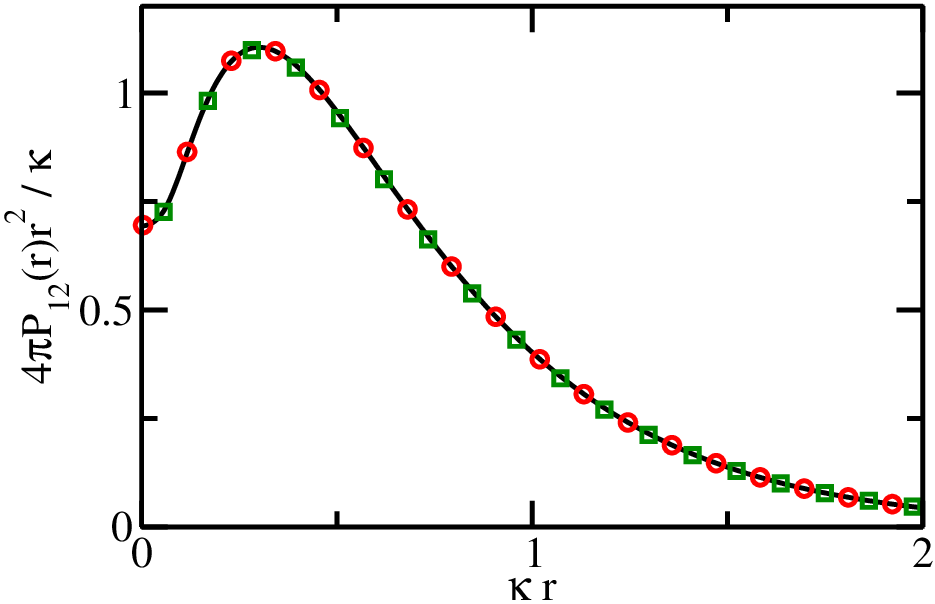}
\caption{(Color online)
  Scaled pair distribution function $4\pi P_{12}(r)r^2$
  for three identical harmonically trapped three-dimensional 
  bosons with two-body
  zero-range interactions with infinitely large $s$-wave scattering length and
  repulsive $1/R^6$ three-body potential.
  The solid line and squares are calculated by the
  zero-temperature PIGS approach and the PIMC approach at $T=\hbar\omega/k_B$.
  For comparison, the circles show the scaled pair
  distribution function 
  obtained by sampling the exact ground state density using the
  Metropolis algorithm.
 }\label{Fig4}
\end{figure}

The squares in Fig.~\ref{Fig4}
show the pair distribution function
calculated by the PIMC approach for three identical bosons at
$T = \hbar \omega / k_B$. For comparison, the solid line and 
the circles show zero-temperature results. The solid line is
calculated using the PIGS approach with a trial wave function that coincides
with the exact ground state wave function~\footnote{
The ground state wave function is obtained by transforming
to hyperspherical coordinates. The hyperangular wave function, which separates
from the hyperradial portion, is known
analytically~\cite{Efimov,BraatenHammerReview} and
the hyperradial wave function is found numerically.
}
while the circles are calculated
by sampling the exact zero-temperature ground state density 
using the Metropolis
algorithm. The agreement between the three sets of calculations
is very good, demonstrating (i) that excitations of the relative
degrees of freedom are negligible at the temperature considered and (ii)
that the path integral approaches accurately resolve the
short-range behavior of the pair distribution function.
The pair distribution functions shown in Fig.~\ref{Fig4} are affected by
the trap, i.e., they move to larger $r$ as the trap frequency
$\omega$ is reduced. The reason is that $\hbar \omega$ is only about four times
smaller than $|E_{\text{trimer}}|$. The magnitude of the $N$-boson energy
$E_{\text{cluster}}$ increases rapidly with
$N$~\cite{stecher10,amy12,kievsky14},
implying that the trap effects decrease quickly with increasing
$N$, thus allowing us to extract the free-space energy 
$E_{\text{cluster}}$ from the finite-temperature trap energies
$E_{\text{sim}}$.

\begin{figure}
\centering
\includegraphics[angle=0,width=0.4\textwidth]{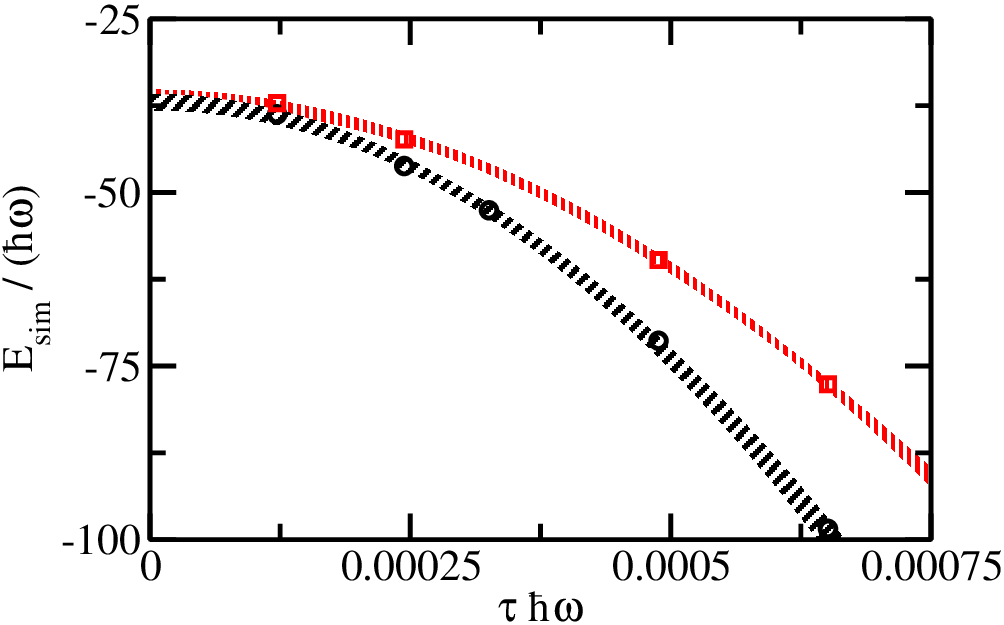}
\caption{(Color online)
  PIMC energies for $N=5$ 
  harmonically trapped three-dimensional bosons with two-body
  zero-range interactions with infinite scattering length and
  and repulsive $1/R^6$ three-body interaction
  at temperature $T=\hbar\omega/k_B$
  as a function of the time step $\tau$.
  The circles (lower-lying data set) and squares (higher-lying data set) 
  show the energy obtained using
  the second- and fourth-order scheme, respectively.
  The error bands are obtained by fitting the data for different $\tau$
  intervals.
 }\label{Fig5}
\end{figure}

Symbols in Fig.~\ref{Fig5}
exemplarily show our PIMC energies $E_{\text{sim}}$
for the five-boson system at $T = \hbar \omega / k_B$ as a 
function of the time step $\tau$. Circles and squares are obtained
using the second- and fourth-order schemes (see earlier discussion),
respectively, to treat the
term $\exp(-\sum_{j<k<l}\tau V_{\text{3b}}(R_{jkl}))$.
The statistical errors are smaller than the symbol size. The fourth-order
results display, as expected, a smaller time step dependence than the
second-order results and are well described by a function of the
form $c_0 +c_2 \tau^2 + c_4 \tau^4$, whereas the second-order results are
described by a function of the form $c_0 +c_2 \tau^2$.
The presence of the $\tau^2$
term for the fourth-order results
is due to the fact that the pair product approximation
neglects three- and higher-body correlations (see also Sec.~\ref{onedsection}).
The shaded regions in Fig.~\ref{Fig5} show errorbands obtained by
fitting the two sets of PIMC energies for different $\tau$ intervals.
The errorbars of the extrapolated
zero time step energies are found to overlap.
We find $E_{\text{sim}}=-37.0(1.2)\hbar\omega$
and $-36.2(1.0)\hbar\omega$
for the second- and fourth-order scheme, respectively.
The free-space energy $E_{\text{cluster}}$ is then obtained by
subtracting the center-of-mass energy,
$E_{\text{cluster}}=E_{\text{sim}}-E_{\text{c.m.}}$.

\begin{figure}
\centering
\includegraphics[angle=0,width=0.4\textwidth]{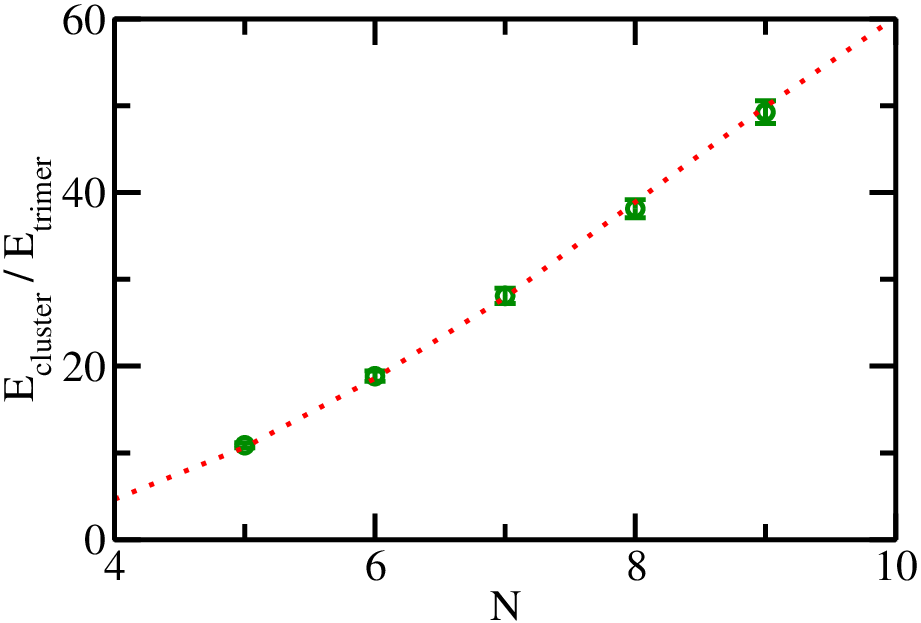}
\caption{(Color online)
  Free-space $N$-boson ground state energy $E_{\text{cluster}}$ 
  as a function of $N$ for infinitely large two-body $s$-wave
  scattering length.
  The circles with errorbars are extracted from our PIMC simulations.
  The dotted line shows the energies reported in Ref.~\cite{stecher10}.
 }\label{Fig6}
\end{figure}

The squares in Fig.~\ref{Fig6}
show $E_{\text{cluster}}$ for $N=5-9$.
The corresponding energies $E_{\text{sim}}$
are obtained using the fourth-order scheme with
$\tau\hbar\omega\approx0.000122$.
As can be seen from Fig.~\ref{Fig5}, this energy lies within the extrapolated
$\tau=0$ errorbands for $N=5$. Figure~\ref{Fig6} scales the energy $E_{\text{cluster}}$
by the corresponding zero-temperature free-space trimer energy $E_{\text{trimer}}$
calculated by the hyperspherical coordinate approach. For comparison,
the dotted line shows the $N$-boson ground state energies from
Ref.~\cite{stecher10} for
a finite-range two-body Gaussian potential with infinitely
large $s$-wave scattering length and a hardcore three-body
regulator. While the model interactions differ, the agreement 
between the two sets of calculations is good, providing further 
support for the (approximate) universality of $N$-boson droplets.
More detailed comparisons will be presented elsewhere~\footnote{
Y. Yan and D. Blume, in preparation.
}.

  \section{conclusion}
  This paper discussed how to treat zero-range two-body interactions in
  $N$-body Monte Carlo simulations. We showed that the incorporation of the
  exact two-body zero-range propagator via the pair product approximation
  allows for an accurate description of paradigmatic strongly-interacting
  one- and three-dimensional model Hamiltonian.

  An important aspect of the studies presented is that the strength of the
  contact interaction requires no renormalization since the simulations are
  performed using the regularized two-body zero-range pseudopotential and 
  continuous spatial coordinates in an unrestricted Hilbert space.
  The fact that the interaction strength does not need to be
  renormalized distinguishes the simulations presented in this paper from
  lattice approaches~\cite{kogut83,lee09,amy13} and
  from configuration interaction
  approaches~\cite{esry99,alhassid08,reimann09}.

  The developments presented in this paper open a number of possibilities. The
  use of two-body zero-range interactions, e.g., provides direct access to the
  two-body Tan contact~\cite{tan1},
  without extrapolation to the zero-range limit. The
  two-body Tan contact is defined for systems with two-body zero-range
  interactions. It relates distinct physical observables such as the large
  momentum tail and aspects of the radio frequency spectrum, and has attracted
  a great deal of theoretical~\cite{Yvan12,platter08,gordon09,timour11,1dtan11,doggen13,contact13}
  and experimental~\cite{jin10,vale10,vale11,jin12,jin14}
  interest. From the computational
  perspective, the adiabatic relation, which involves the change of the energy
  with the scattering length, and the pair relation~\cite{tan1,daily09},
  which gives the
  probability of finding two particles at the same position, are most
  convenient. Earlier work applied these relations to systems with
  finite-range interactions and extrapolated to the
  zero-range limit. Using our
  zero-range propagators, these relations can be used directly for the
  determination of the Tan contact, eliminating the extrapolation step.

{\em{Acknowledgement:}}
We thank Ebrahim Gharashi for providing the energies of three identical bosons
calculated using the approach of Ref.~\cite{ebrahim12} and
Maurizio Rossi for very useful comments on an earlier
version of the manuscript.
Support by the National
Science Foundation (NSF) through Grant No.
PHY-1415112
is gratefully acknowledged.
This work used the Extreme Science and Engineering
Discovery Environment (XSEDE), which is supported by
NSF Grant No. OCI-1053575, and the
WSU HPC.


\begin{thebibliography}{80}%
\makeatletter
\providecommand \@ifxundefined [1]{%
 \@ifx{#1\undefined}
}%
\providecommand \@ifnum [1]{%
 \ifnum #1\expandafter \@firstoftwo
 \else \expandafter \@secondoftwo
 \fi
}%
\providecommand \@ifx [1]{%
 \ifx #1\expandafter \@firstoftwo
 \else \expandafter \@secondoftwo
 \fi
}%
\providecommand \natexlab [1]{#1}%
\providecommand \enquote  [1]{``#1''}%
\providecommand \bibnamefont  [1]{#1}%
\providecommand \bibfnamefont [1]{#1}%
\providecommand \citenamefont [1]{#1}%
\providecommand \href@noop [0]{\@secondoftwo}%
\providecommand \href [0]{\begingroup \@sanitize@url \@href}%
\providecommand \@href[1]{\@@startlink{#1}\@@href}%
\providecommand \@@href[1]{\endgroup#1\@@endlink}%
\providecommand \@sanitize@url [0]{\catcode `\\12\catcode `\$12\catcode
  `\&12\catcode `\#12\catcode `\^12\catcode `\_12\catcode `\%12\relax}%
\providecommand \@@startlink[1]{}%
\providecommand \@@endlink[0]{}%
\providecommand \url  [0]{\begingroup\@sanitize@url \@url }%
\providecommand \@url [1]{\endgroup\@href {#1}{\urlprefix }}%
\providecommand \urlprefix  [0]{URL }%
\providecommand \Eprint [0]{\href }%
\providecommand \doibase [0]{http://dx.doi.org/}%
\providecommand \selectlanguage [0]{\@gobble}%
\providecommand \bibinfo  [0]{\@secondoftwo}%
\providecommand \bibfield  [0]{\@secondoftwo}%
\providecommand \translation [1]{[#1]}%
\providecommand \BibitemOpen [0]{}%
\providecommand \bibitemStop [0]{}%
\providecommand \bibitemNoStop [0]{.\EOS\space}%
\providecommand \EOS [0]{\spacefactor3000\relax}%
\providecommand \BibitemShut  [1]{\csname bibitem#1\endcsname}%
\let\auto@bib@innerbib\@empty
\bibitem [{\citenamefont {Fermi}(1934)}]{fermi34}%
  \BibitemOpen
  \bibfield  {author} {\bibinfo {author} {\bibfnamefont {E.}~\bibnamefont
  {Fermi}},\ }\href@noop {} {\bibfield  {journal} {\bibinfo  {journal} {Nuovo
  Cimento}\ }\textbf {\bibinfo {volume} {11}},\ \bibinfo {pages} {157}
  (\bibinfo {year} {1934})}\BibitemShut {NoStop}%
\bibitem [{\citenamefont {Claude Cohen-Tannoudji}()}]{claudequantum}%
  \BibitemOpen
  \bibfield  {author} {\bibinfo {author} {\bibfnamefont {B.}~\bibnamefont
  {Claude Cohen-Tannoudji}},\ }\href@noop {} {\emph {\bibinfo {title} {Quantum
  Mechanics}}},\ \bibinfo {number} {vol. 1}\ (\bibinfo  {publisher} {Hermann,
  Paris})\BibitemShut {NoStop}%
\bibitem [{\citenamefont {Demkov}\ and\ \citenamefont
  {Ostrovski{\u\i}}(1988)}]{demkov88}%
  \BibitemOpen
  \bibfield  {author} {\bibinfo {author} {\bibfnamefont {Y.~N.}\ \bibnamefont
  {Demkov}}\ and\ \bibinfo {author} {\bibfnamefont {V.~N.}\ \bibnamefont
  {Ostrovski{\u\i}}},\ }\href@noop {} {\emph {\bibinfo {title} {Zero-Range
  Potentials and Their Applications in Atomic Physics}}}\ (\bibinfo
  {publisher} {Plenum Press, New York},\ \bibinfo {year} {1988})\BibitemShut
  {NoStop}%
\bibitem [{\citenamefont {Krsti\ifmmode~\acute{c}\else \'{c}\fi{}}\ \emph
  {et~al.}(1991)\citenamefont {Krsti\ifmmode~\acute{c}\else \'{c}\fi{}},
  \citenamefont {Milo\v{s}evi\ifmmode~\acute{c}\else \'{c}\fi{}},\ and\
  \citenamefont {Janev}}]{janev91}%
  \BibitemOpen
  \bibfield  {author} {\bibinfo {author} {\bibfnamefont {P.~S.}\ \bibnamefont
  {Krsti\ifmmode~\acute{c}\else \'{c}\fi{}}}, \bibinfo {author} {\bibfnamefont
  {D.~B.}\ \bibnamefont {Milo\v{s}evi\ifmmode~\acute{c}\else \'{c}\fi{}}}, \
  and\ \bibinfo {author} {\bibfnamefont {R.~K.}\ \bibnamefont {Janev}},\ }\href
  {\doibase 10.1103/PhysRevA.44.3089} {\bibfield  {journal} {\bibinfo
  {journal} {Phys. Rev. A}\ }\textbf {\bibinfo {volume} {44}},\ \bibinfo
  {pages} {3089} (\bibinfo {year} {1991})}\BibitemShut {NoStop}%
\bibitem [{\citenamefont {Manakov}\ \emph {et~al.}(2000)\citenamefont
  {Manakov}, \citenamefont {Frolov}, \citenamefont {Starace},\ and\
  \citenamefont {Fabrikant}}]{fabrikant00}%
  \BibitemOpen
  \bibfield  {author} {\bibinfo {author} {\bibfnamefont {N.~L.}\ \bibnamefont
  {Manakov}}, \bibinfo {author} {\bibfnamefont {M.~V.}\ \bibnamefont {Frolov}},
  \bibinfo {author} {\bibfnamefont {A.~F.}\ \bibnamefont {Starace}}, \ and\
  \bibinfo {author} {\bibfnamefont {I.~I.}\ \bibnamefont {Fabrikant}},\ }\href
  {http://stacks.iop.org/0953-4075/33/i=15/a=201} {\bibfield  {journal}
  {\bibinfo  {journal} {J. Phys. B}\ }\textbf {\bibinfo {volume} {33}},\
  \bibinfo {pages} {R141} (\bibinfo {year} {2000})}\BibitemShut {NoStop}%
\bibitem [{\citenamefont {Girardeau}(1960)}]{girardeau60}%
  \BibitemOpen
  \bibfield  {author} {\bibinfo {author} {\bibfnamefont {M.}~\bibnamefont
  {Girardeau}},\ }\href@noop {} {\bibfield  {journal} {\bibinfo  {journal} {J
  Math. Phys.}\ }\textbf {\bibinfo {volume} {1}} (\bibinfo {year}
  {1960})}\BibitemShut {NoStop}%
\bibitem [{\citenamefont {Lieb}\ and\ \citenamefont {Liniger}(1963)}]{lieb63}%
  \BibitemOpen
  \bibfield  {author} {\bibinfo {author} {\bibfnamefont {E.~H.}\ \bibnamefont
  {Lieb}}\ and\ \bibinfo {author} {\bibfnamefont {W.}~\bibnamefont {Liniger}},\
  }\href {\doibase 10.1103/PhysRev.130.1605} {\bibfield  {journal} {\bibinfo
  {journal} {Phys. Rev.}\ }\textbf {\bibinfo {volume} {130}},\ \bibinfo {pages}
  {1605} (\bibinfo {year} {1963})}\BibitemShut {NoStop}%
\bibitem [{\citenamefont {{Gaudin}}(1967)}]{gaudin67}%
  \BibitemOpen
  \bibfield  {author} {\bibinfo {author} {\bibfnamefont {M.}~\bibnamefont
  {{Gaudin}}},\ }\href {\doibase 10.1016/0375-9601(67)90193-4} {\bibfield
  {journal} {\bibinfo  {journal} {Phys. Lett. A}\ }\textbf {\bibinfo {volume}
  {24}},\ \bibinfo {pages} {55} (\bibinfo {year} {1967})}\BibitemShut {NoStop}%
\bibitem [{\citenamefont {{Yang}}(1967)}]{yang67}%
  \BibitemOpen
  \bibfield  {author} {\bibinfo {author} {\bibfnamefont {C.~N.}\ \bibnamefont
  {{Yang}}},\ }\href {\doibase 10.1103/PhysRevLett.19.1312} {\bibfield
  {journal} {\bibinfo  {journal} {Phys. Rev. Lett.}\ }\textbf {\bibinfo
  {volume} {19}},\ \bibinfo {pages} {1312} (\bibinfo {year}
  {1967})}\BibitemShut {NoStop}%
\bibitem [{\citenamefont {Yang}\ and\ \citenamefont {Yang}(1969)}]{yang69}%
  \BibitemOpen
  \bibfield  {author} {\bibinfo {author} {\bibfnamefont {C.~N.}\ \bibnamefont
  {Yang}}\ and\ \bibinfo {author} {\bibfnamefont {C.~P.}\ \bibnamefont
  {Yang}},\ }\href@noop {} {\bibfield  {journal} {\bibinfo  {journal} {J. Math.
  Phys.}\ }\textbf {\bibinfo {volume} {10}} (\bibinfo {year}
  {1969})}\BibitemShut {NoStop}%
\bibitem [{\citenamefont {Jones}\ and\ \citenamefont
  {March}(1985)}]{jones1985theoretical}%
  \BibitemOpen
  \bibfield  {author} {\bibinfo {author} {\bibfnamefont {W.}~\bibnamefont
  {Jones}}\ and\ \bibinfo {author} {\bibfnamefont {N.~H.}\ \bibnamefont
  {March}},\ }\href@noop {} {\emph {\bibinfo {title} {Theoretical Solid State
  Physics}}},\ \bibinfo {number} {vol. 2}\ (\bibinfo  {publisher} {Dover
  Publications, New York},\ \bibinfo {year} {1985})\BibitemShut {NoStop}%
\bibitem [{\citenamefont {Bloch}\ \emph {et~al.}(2008)\citenamefont {Bloch},
  \citenamefont {Dalibard},\ and\ \citenamefont {Zwerger}}]{blochreview}%
  \BibitemOpen
  \bibfield  {author} {\bibinfo {author} {\bibfnamefont {I.}~\bibnamefont
  {Bloch}}, \bibinfo {author} {\bibfnamefont {J.}~\bibnamefont {Dalibard}}, \
  and\ \bibinfo {author} {\bibfnamefont {W.}~\bibnamefont {Zwerger}},\ }\href
  {\doibase 10.1103/RevModPhys.80.885} {\bibfield  {journal} {\bibinfo
  {journal} {Rev. Mod. Phys.}\ }\textbf {\bibinfo {volume} {80}},\ \bibinfo
  {pages} {885} (\bibinfo {year} {2008})}\BibitemShut {NoStop}%
\bibitem [{\citenamefont {Dalfovo}\ \emph {et~al.}(1999)\citenamefont
  {Dalfovo}, \citenamefont {Giorgini}, \citenamefont {Pitaevskii},\ and\
  \citenamefont {Stringari}}]{stringariBosereview}%
  \BibitemOpen
  \bibfield  {author} {\bibinfo {author} {\bibfnamefont {F.}~\bibnamefont
  {Dalfovo}}, \bibinfo {author} {\bibfnamefont {S.}~\bibnamefont {Giorgini}},
  \bibinfo {author} {\bibfnamefont {L.~P.}\ \bibnamefont {Pitaevskii}}, \ and\
  \bibinfo {author} {\bibfnamefont {S.}~\bibnamefont {Stringari}},\ }\href
  {\doibase 10.1103/RevModPhys.71.463} {\bibfield  {journal} {\bibinfo
  {journal} {Rev. Mod. Phys.}\ }\textbf {\bibinfo {volume} {71}},\ \bibinfo
  {pages} {463} (\bibinfo {year} {1999})}\BibitemShut {NoStop}%
\bibitem [{\citenamefont {Giorgini}\ \emph {et~al.}(2008)\citenamefont
  {Giorgini}, \citenamefont {Pitaevskii},\ and\ \citenamefont
  {Stringari}}]{stringariFermireview}%
  \BibitemOpen
  \bibfield  {author} {\bibinfo {author} {\bibfnamefont {S.}~\bibnamefont
  {Giorgini}}, \bibinfo {author} {\bibfnamefont {L.~P.}\ \bibnamefont
  {Pitaevskii}}, \ and\ \bibinfo {author} {\bibfnamefont {S.}~\bibnamefont
  {Stringari}},\ }\href {\doibase 10.1103/RevModPhys.80.1215} {\bibfield
  {journal} {\bibinfo  {journal} {Rev. Mod. Phys.}\ }\textbf {\bibinfo {volume}
  {80}},\ \bibinfo {pages} {1215} (\bibinfo {year} {2008})}\BibitemShut
  {NoStop}%
\bibitem [{\citenamefont {Efimov}(1970)}]{Efimov}%
  \BibitemOpen
  \bibfield  {author} {\bibinfo {author} {\bibfnamefont {V.}~\bibnamefont
  {Efimov}},\ }\href@noop {} {\bibfield  {journal} {\bibinfo  {journal} {Phys.
  Lett. B}\ }\textbf {\bibinfo {volume} {33}},\ \bibinfo {pages} {563}
  (\bibinfo {year} {1970})}\BibitemShut {NoStop}%
\bibitem [{\citenamefont {Fedorov}\ \emph {et~al.}(1994)\citenamefont
  {Fedorov}, \citenamefont {Jensen},\ and\ \citenamefont
  {Riisager}}]{fedorov94}%
  \BibitemOpen
  \bibfield  {author} {\bibinfo {author} {\bibfnamefont {D.~V.}\ \bibnamefont
  {Fedorov}}, \bibinfo {author} {\bibfnamefont {A.~S.}\ \bibnamefont {Jensen}},
  \ and\ \bibinfo {author} {\bibfnamefont {K.}~\bibnamefont {Riisager}},\
  }\href {\doibase 10.1103/PhysRevC.49.201} {\bibfield  {journal} {\bibinfo
  {journal} {Phys. Rev. C}\ }\textbf {\bibinfo {volume} {49}},\ \bibinfo
  {pages} {201} (\bibinfo {year} {1994})}\BibitemShut {NoStop}%
\bibitem [{\citenamefont {Braaten}\ and\ \citenamefont
  {Hammer}(2006)}]{BraatenHammerReview}%
  \BibitemOpen
  \bibfield  {author} {\bibinfo {author} {\bibfnamefont {E.}~\bibnamefont
  {Braaten}}\ and\ \bibinfo {author} {\bibfnamefont {H.-W.}\ \bibnamefont
  {Hammer}},\ }\href {\doibase 10.1016/j.physrep.2006.03.001} {\bibfield
  {journal} {\bibinfo  {journal} {Phys. Rep.}\ }\textbf {\bibinfo {volume}
  {428}},\ \bibinfo {pages} {259 } (\bibinfo {year} {2006})}\BibitemShut
  {NoStop}%
\bibitem [{\citenamefont {{Ceperley}}(1995)}]{ceperleyrev}%
  \BibitemOpen
  \bibfield  {author} {\bibinfo {author} {\bibfnamefont {D.~M.}\ \bibnamefont
  {{Ceperley}}},\ }\href {\doibase 10.1103/RevModPhys.67.279} {\bibfield
  {journal} {\bibinfo  {journal} {Rev. Mod. Phys.}\ }\textbf {\bibinfo {volume}
  {67}},\ \bibinfo {pages} {279} (\bibinfo {year} {1995})}\BibitemShut
  {NoStop}%
\bibitem [{\citenamefont {Boninsegni}(2005)}]{boninsegni05}%
  \BibitemOpen
  \bibfield  {author} {\bibinfo {author} {\bibfnamefont {M.}~\bibnamefont
  {Boninsegni}},\ }\href {\doibase 10.1007/s10909-005-7513-0} {\bibfield
  {journal} {\bibinfo  {journal} {J. Low Temp. Phys.}\ }\textbf {\bibinfo
  {volume} {141}},\ \bibinfo {pages} {27} (\bibinfo {year} {2005})}\BibitemShut
  {NoStop}%
\bibitem [{\citenamefont {Krauth}(2006)}]{krauth2006statistical}%
  \BibitemOpen
  \bibfield  {author} {\bibinfo {author} {\bibfnamefont {W.}~\bibnamefont
  {Krauth}},\ }\href@noop {} {\emph {\bibinfo {title} {Statistical Mechanics:
  Algorithms and Computations}}},\ Oxford Master Series in Physics\ (\bibinfo
  {publisher} {Oxford University Press, Oxford, UK},\ \bibinfo {year}
  {2006})\BibitemShut {NoStop}%
\bibitem [{\citenamefont {Het{\'e}nyi}\ \emph {et~al.}(1999)\citenamefont
  {Het{\'e}nyi}, \citenamefont {Rabani},\ and\ \citenamefont {Berne}}]{pigs99}%
  \BibitemOpen
  \bibfield  {author} {\bibinfo {author} {\bibfnamefont {B.}~\bibnamefont
  {Het{\'e}nyi}}, \bibinfo {author} {\bibfnamefont {E.}~\bibnamefont {Rabani}},
  \ and\ \bibinfo {author} {\bibfnamefont {B.~J.}\ \bibnamefont {Berne}},\
  }\href {\doibase 10.1063/1.478520} {\bibfield  {journal} {\bibinfo  {journal}
  {J. Chem. Phys.}\ }\textbf {\bibinfo {volume} {110}},\ \bibinfo {pages}
  {6143} (\bibinfo {year} {1999})}\BibitemShut {NoStop}%
\bibitem [{\citenamefont {Sarsa}\ \emph {et~al.}(2000)\citenamefont {Sarsa},
  \citenamefont {Schmidt},\ and\ \citenamefont {Magro}}]{pigs00}%
  \BibitemOpen
  \bibfield  {author} {\bibinfo {author} {\bibfnamefont {A.}~\bibnamefont
  {Sarsa}}, \bibinfo {author} {\bibfnamefont {K.~E.}\ \bibnamefont {Schmidt}},
  \ and\ \bibinfo {author} {\bibfnamefont {W.~R.}\ \bibnamefont {Magro}},\
  }\href {\doibase 10.1063/1.481926} {\bibfield  {journal} {\bibinfo  {journal}
  {J. Chem. Phys.}\ }\textbf {\bibinfo {volume} {113}},\ \bibinfo {pages}
  {1366} (\bibinfo {year} {2000})}\BibitemShut {NoStop}%
\bibitem [{\citenamefont {Cuervo}\ \emph {et~al.}(2005)\citenamefont {Cuervo},
  \citenamefont {Roy},\ and\ \citenamefont {Boninsegni}}]{massimo05}%
  \BibitemOpen
  \bibfield  {author} {\bibinfo {author} {\bibfnamefont {J.~E.}\ \bibnamefont
  {Cuervo}}, \bibinfo {author} {\bibfnamefont {P.-N.}\ \bibnamefont {Roy}}, \
  and\ \bibinfo {author} {\bibfnamefont {M.}~\bibnamefont {Boninsegni}},\
  }\href {\doibase 10.1063/1.1872775} {\bibfield  {journal} {\bibinfo
  {journal} {J. Chem. Phys.}\ }\textbf {\bibinfo {volume} {122}},\ \bibinfo
  {eid} {114504} (\bibinfo {year} {2005})}\BibitemShut {NoStop}%
\bibitem [{\citenamefont {Rossi}\ \emph {et~al.}(2009)\citenamefont {Rossi},
  \citenamefont {Nava}, \citenamefont {Reatto},\ and\ \citenamefont
  {Galli}}]{pigs08}%
  \BibitemOpen
  \bibfield  {author} {\bibinfo {author} {\bibfnamefont {M.}~\bibnamefont
  {Rossi}}, \bibinfo {author} {\bibfnamefont {M.}~\bibnamefont {Nava}},
  \bibinfo {author} {\bibfnamefont {L.}~\bibnamefont {Reatto}}, \ and\ \bibinfo
  {author} {\bibfnamefont {D.~E.}\ \bibnamefont {Galli}},\ }\href {\doibase
  http://dx.doi.org/10.1063/1.3247833} {\bibfield  {journal} {\bibinfo
  {journal} {J. Chem. Phys.}\ }\textbf {\bibinfo {volume} {131}},\ \bibinfo
  {eid} {154108} (\bibinfo {year} {2009})}\BibitemShut {NoStop}%
\bibitem [{\citenamefont {Gaveau}\ and\ \citenamefont
  {Schulman}(1986)}]{schulman86}%
  \BibitemOpen
  \bibfield  {author} {\bibinfo {author} {\bibfnamefont {B.}~\bibnamefont
  {Gaveau}}\ and\ \bibinfo {author} {\bibfnamefont {L.~S.}\ \bibnamefont
  {Schulman}},\ }\href {http://stacks.iop.org/0305-4470/19/i=10/a=024}
  {\bibfield  {journal} {\bibinfo  {journal} {J. Phys. A}\ }\textbf {\bibinfo
  {volume} {19}},\ \bibinfo {pages} {1833} (\bibinfo {year}
  {1986})}\BibitemShut {NoStop}%
\bibitem [{\citenamefont {Blinder}(1988)}]{blinder88}%
  \BibitemOpen
  \bibfield  {author} {\bibinfo {author} {\bibfnamefont {S.~M.}\ \bibnamefont
  {Blinder}},\ }\href {\doibase 10.1103/PhysRevA.37.973} {\bibfield  {journal}
  {\bibinfo  {journal} {Phys. Rev. A}\ }\textbf {\bibinfo {volume} {37}},\
  \bibinfo {pages} {973} (\bibinfo {year} {1988})}\BibitemShut {NoStop}%
\bibitem [{\citenamefont {Lawande}\ and\ \citenamefont
  {Bhagwat}(1988)}]{lawande88}%
  \BibitemOpen
  \bibfield  {author} {\bibinfo {author} {\bibfnamefont {S.~V.}\ \bibnamefont
  {Lawande}}\ and\ \bibinfo {author} {\bibfnamefont {K.~V.}\ \bibnamefont
  {Bhagwat}},\ }\href {\doibase 10.1016/0375-9601(88)90622-6} {\bibfield
  {journal} {\bibinfo  {journal} {Phys. Lett. A}\ }\textbf {\bibinfo {volume}
  {131}},\ \bibinfo {pages} {8 } (\bibinfo {year} {1988})}\BibitemShut
  {NoStop}%
\bibitem [{\citenamefont {W\'odkiewicz}(1991)}]{wodkiewicz91}%
  \BibitemOpen
  \bibfield  {author} {\bibinfo {author} {\bibfnamefont {K.}~\bibnamefont
  {W\'odkiewicz}},\ }\href {\doibase 10.1103/PhysRevA.43.68} {\bibfield
  {journal} {\bibinfo  {journal} {Phys. Rev. A}\ }\textbf {\bibinfo {volume}
  {43}},\ \bibinfo {pages} {68} (\bibinfo {year} {1991})}\BibitemShut {NoStop}%
\bibitem [{\citenamefont {Pathria}(1996)}]{pathria1996statistical}%
  \BibitemOpen
  \bibfield  {author} {\bibinfo {author} {\bibfnamefont {R.}~\bibnamefont
  {Pathria}},\ }\href@noop {} {\emph {\bibinfo {title} {Statistical
  Mechanics}}}\ (\bibinfo  {publisher} {Elsevier, Amsterdam},\ \bibinfo {year}
  {1996})\BibitemShut {NoStop}%
\bibitem [{\citenamefont {Huang}(1987)}]{huang1987statistical}%
  \BibitemOpen
  \bibfield  {author} {\bibinfo {author} {\bibfnamefont {K.}~\bibnamefont
  {Huang}},\ }\href@noop {} {\emph {\bibinfo {title} {Statistical
  Mechanics}}},\ \bibinfo {edition} {2nd}\ ed.\ (\bibinfo  {publisher} {Wiley,
  New York},\ \bibinfo {year} {1987})\BibitemShut {NoStop}%
\bibitem [{Note1()}]{Note1}%
  \BibitemOpen
  \bibinfo {note} {Equation~(\ref {normcondition}) is to be interpreted as
  follows: $\DOTSI \intop \ilimits@ \protect \tmspace -\thinmuskip
  {.1667em}[\psi _{k}^s(x)]^*\psi _{k'}^s(x)\protect \tmspace +\thinmuskip
  {.1667em}\protect \mathrm {d}x=\delta (k-k'),$ $\DOTSI \intop \ilimits@
  \protect \tmspace -\thinmuskip {.1667em}[\psi _{k}^a(x)]^*\psi
  _{k'}^a(x)\protect \tmspace +\thinmuskip {.1667em}\protect \mathrm
  {d}x=\delta (k-k'),$ and $\DOTSI \intop \ilimits@ \protect \tmspace
  -\thinmuskip {.1667em}[\psi _{k}^a(x)]^*\psi _{k'}^s(x)\protect \tmspace
  +\thinmuskip {.1667em}\protect \mathrm {d}x=0$ (due to
  symmetry).}\BibitemShut {Stop}%
\bibitem [{\citenamefont {Casula}\ \emph {et~al.}(2008)\citenamefont {Casula},
  \citenamefont {Ceperley},\ and\ \citenamefont {Mueller}}]{ceperley08}%
  \BibitemOpen
  \bibfield  {author} {\bibinfo {author} {\bibfnamefont {M.}~\bibnamefont
  {Casula}}, \bibinfo {author} {\bibfnamefont {D.~M.}\ \bibnamefont
  {Ceperley}}, \ and\ \bibinfo {author} {\bibfnamefont {E.~J.}\ \bibnamefont
  {Mueller}},\ }\href {\doibase 10.1103/PhysRevA.78.033607} {\bibfield
  {journal} {\bibinfo  {journal} {Phys. Rev. A}\ }\textbf {\bibinfo {volume}
  {78}},\ \bibinfo {pages} {033607} (\bibinfo {year} {2008})}\BibitemShut
  {NoStop}%
\bibitem [{\citenamefont {Olshanii}(1998)}]{olshanii98}%
  \BibitemOpen
  \bibfield  {author} {\bibinfo {author} {\bibfnamefont {M.}~\bibnamefont
  {Olshanii}},\ }\href {\doibase 10.1103/PhysRevLett.81.938} {\bibfield
  {journal} {\bibinfo  {journal} {Phys. Rev. Lett.}\ }\textbf {\bibinfo
  {volume} {81}},\ \bibinfo {pages} {938} (\bibinfo {year} {1998})}\BibitemShut
  {NoStop}%
\bibitem [{\citenamefont {Girardeau}\ \emph {et~al.}(2004)\citenamefont
  {Girardeau}, \citenamefont {Nguyen},\ and\ \citenamefont
  {Olshanii}}]{girardeau04}%
  \BibitemOpen
  \bibfield  {author} {\bibinfo {author} {\bibfnamefont {M.~D.}\ \bibnamefont
  {Girardeau}}, \bibinfo {author} {\bibfnamefont {H.}~\bibnamefont {Nguyen}}, \
  and\ \bibinfo {author} {\bibfnamefont {M.}~\bibnamefont {Olshanii}},\ }\href
  {\doibase 10.1016/j.optcom.2004.09.079} {\bibfield  {journal} {\bibinfo
  {journal} {Opt. Commun.}\ }\textbf {\bibinfo {volume} {243}},\ \bibinfo
  {pages} {3 } (\bibinfo {year} {2004})}\BibitemShut {NoStop}%
\bibitem [{\citenamefont {Busch}\ \emph {et~al.}(1998)\citenamefont {Busch},
  \citenamefont {Englert}, \citenamefont {Rza{\.z}ewski},\ and\ \citenamefont
  {Wilkens}}]{busch}%
  \BibitemOpen
  \bibfield  {author} {\bibinfo {author} {\bibfnamefont {T.}~\bibnamefont
  {Busch}}, \bibinfo {author} {\bibfnamefont {B.-G.}\ \bibnamefont {Englert}},
  \bibinfo {author} {\bibfnamefont {K.}~\bibnamefont {Rza{\.z}ewski}}, \ and\
  \bibinfo {author} {\bibfnamefont {M.}~\bibnamefont {Wilkens}},\ }\href
  {\doibase 10.1023/A:1018705520999} {\bibfield  {journal} {\bibinfo  {journal}
  {Found. Phys.}\ }\textbf {\bibinfo {volume} {28}},\ \bibinfo {pages} {549}
  (\bibinfo {year} {1998})}\BibitemShut {NoStop}%
\bibitem [{\citenamefont {Trotter}(1959)}]{trotter1959product}%
  \BibitemOpen
  \bibfield  {author} {\bibinfo {author} {\bibfnamefont {H.~F.}\ \bibnamefont
  {Trotter}},\ }\href@noop {} {\bibfield  {journal} {\bibinfo  {journal} {Proc.
  Amer. Math. Soc.}\ }\textbf {\bibinfo {volume} {10}},\ \bibinfo {pages} {545}
  (\bibinfo {year} {1959})}\BibitemShut {NoStop}%
\bibitem [{\citenamefont {Huang}\ and\ \citenamefont
  {Yang}(1957)}]{zerorangepotential}%
  \BibitemOpen
  \bibfield  {author} {\bibinfo {author} {\bibfnamefont {K.}~\bibnamefont
  {Huang}}\ and\ \bibinfo {author} {\bibfnamefont {C.~N.}\ \bibnamefont
  {Yang}},\ }\href {\doibase 10.1103/PhysRev.105.767} {\bibfield  {journal}
  {\bibinfo  {journal} {Phys. Rev.}\ }\textbf {\bibinfo {volume} {105}},\
  \bibinfo {pages} {767} (\bibinfo {year} {1957})}\BibitemShut {NoStop}%
\bibitem [{Note2()}]{Note2}%
  \BibitemOpen
  \bibinfo {note} {See entry 3.954 of I. S. Gradshteyn and I. M. Ryzhik,
  {\protect \em {Table of Integrals, Series, and Products}}, 6th Ed., Academic
  Press.}\BibitemShut {Stop}%
\bibitem [{\citenamefont {{Pessoa}}\ \emph {et~al.}()\citenamefont {{Pessoa}},
  \citenamefont {{Vitiello}},\ and\ \citenamefont {{Schmidt}}}]{schmidt14}%
  \BibitemOpen
  \bibfield  {author} {\bibinfo {author} {\bibfnamefont {R.}~\bibnamefont
  {{Pessoa}}}, \bibinfo {author} {\bibfnamefont {S.~A.}\ \bibnamefont
  {{Vitiello}}}, \ and\ \bibinfo {author} {\bibfnamefont {K.~E.}\ \bibnamefont
  {{Schmidt}}},\ }\href@noop {} {\ }\Eprint {http://arxiv.org/abs/1411.5960}
  {arXiv:1411.5960} \BibitemShut {NoStop}%
\bibitem [{\citenamefont {{Gharashi}}\ \emph {et~al.}(2012)\citenamefont
  {{Gharashi}}, \citenamefont {{Daily}},\ and\ \citenamefont
  {{Blume}}}]{ebrahim12}%
  \BibitemOpen
  \bibfield  {author} {\bibinfo {author} {\bibfnamefont {S.~E.}\ \bibnamefont
  {{Gharashi}}}, \bibinfo {author} {\bibfnamefont {K.~M.}\ \bibnamefont
  {{Daily}}}, \ and\ \bibinfo {author} {\bibfnamefont {D.}~\bibnamefont
  {{Blume}}},\ }\href {\doibase 10.1103/PhysRevA.86.042702} {\bibfield
  {journal} {\bibinfo  {journal} {Phys. Rev. A}\ }\textbf {\bibinfo {volume}
  {86}},\ \bibinfo {eid} {042702} (\bibinfo {year} {2012})}\BibitemShut
  {NoStop}%
\bibitem [{\citenamefont {{Barth}}\ and\ \citenamefont
  {{Zwerger}}(2011)}]{1dtan11}%
  \BibitemOpen
  \bibfield  {author} {\bibinfo {author} {\bibfnamefont {M.}~\bibnamefont
  {{Barth}}}\ and\ \bibinfo {author} {\bibfnamefont {W.}~\bibnamefont
  {{Zwerger}}},\ }\href {\doibase 10.1016/j.aop.2011.05.010} {\bibfield
  {journal} {\bibinfo  {journal} {Ann. Phys. (NY)}\ }\textbf {\bibinfo {volume}
  {326}},\ \bibinfo {pages} {2544} (\bibinfo {year} {2011})}\BibitemShut
  {NoStop}%
\bibitem [{\citenamefont {{Tan}}(2008)}]{tan1}%
  \BibitemOpen
  \bibfield  {author} {\bibinfo {author} {\bibfnamefont {S.}~\bibnamefont
  {{Tan}}},\ }\href {\doibase 10.1016/j.aop.2008.03.004} {\bibfield  {journal}
  {\bibinfo  {journal} {Ann. Phys. (NY)}\ }\textbf {\bibinfo {volume} {323}},\
  \bibinfo {pages} {2952} (\bibinfo {year} {2008})}\BibitemShut {NoStop}%
\bibitem [{\citenamefont {{Piatecki}}\ and\ \citenamefont
  {{Krauth}}(2014)}]{krauth13}%
  \BibitemOpen
  \bibfield  {author} {\bibinfo {author} {\bibfnamefont {S.}~\bibnamefont
  {{Piatecki}}}\ and\ \bibinfo {author} {\bibfnamefont {W.}~\bibnamefont
  {{Krauth}}},\ }\href {\doibase 10.1038/ncomms4503} {\bibfield  {journal}
  {\bibinfo  {journal} {Nat. Commun.}\ }\textbf {\bibinfo {volume} {5}},\
  \bibinfo {pages} {3503} (\bibinfo {year} {2014})}\BibitemShut {NoStop}%
\bibitem [{\citenamefont {Rem}\ \emph {et~al.}(2013)\citenamefont {Rem},
  \citenamefont {Grier}, \citenamefont {Ferrier-Barbut}, \citenamefont
  {Eismann}, \citenamefont {Langen}, \citenamefont {Navon}, \citenamefont
  {Khaykovich}, \citenamefont {Werner}, \citenamefont {Petrov}, \citenamefont
  {Chevy},\ and\ \citenamefont {Salomon}}]{salomon13}%
  \BibitemOpen
  \bibfield  {author} {\bibinfo {author} {\bibfnamefont {B.~S.}\ \bibnamefont
  {Rem}}, \bibinfo {author} {\bibfnamefont {A.~T.}\ \bibnamefont {Grier}},
  \bibinfo {author} {\bibfnamefont {I.}~\bibnamefont {Ferrier-Barbut}},
  \bibinfo {author} {\bibfnamefont {U.}~\bibnamefont {Eismann}}, \bibinfo
  {author} {\bibfnamefont {T.}~\bibnamefont {Langen}}, \bibinfo {author}
  {\bibfnamefont {N.}~\bibnamefont {Navon}}, \bibinfo {author} {\bibfnamefont
  {L.}~\bibnamefont {Khaykovich}}, \bibinfo {author} {\bibfnamefont
  {F.}~\bibnamefont {Werner}}, \bibinfo {author} {\bibfnamefont {D.~S.}\
  \bibnamefont {Petrov}}, \bibinfo {author} {\bibfnamefont {F.}~\bibnamefont
  {Chevy}}, \ and\ \bibinfo {author} {\bibfnamefont {C.}~\bibnamefont
  {Salomon}},\ }\href {\doibase 10.1103/PhysRevLett.110.163202} {\bibfield
  {journal} {\bibinfo  {journal} {Phys. Rev. Lett.}\ }\textbf {\bibinfo
  {volume} {110}},\ \bibinfo {pages} {163202} (\bibinfo {year}
  {2013})}\BibitemShut {NoStop}%
\bibitem [{\citenamefont {Fletcher}\ \emph {et~al.}(2013)\citenamefont
  {Fletcher}, \citenamefont {Gaunt}, \citenamefont {Navon}, \citenamefont
  {Smith},\ and\ \citenamefont {Hadzibabic}}]{hadzibabic13}%
  \BibitemOpen
  \bibfield  {author} {\bibinfo {author} {\bibfnamefont {R.~J.}\ \bibnamefont
  {Fletcher}}, \bibinfo {author} {\bibfnamefont {A.~L.}\ \bibnamefont {Gaunt}},
  \bibinfo {author} {\bibfnamefont {N.}~\bibnamefont {Navon}}, \bibinfo
  {author} {\bibfnamefont {R.~P.}\ \bibnamefont {Smith}}, \ and\ \bibinfo
  {author} {\bibfnamefont {Z.}~\bibnamefont {Hadzibabic}},\ }\href {\doibase
  10.1103/PhysRevLett.111.125303} {\bibfield  {journal} {\bibinfo  {journal}
  {Phys. Rev. Lett.}\ }\textbf {\bibinfo {volume} {111}},\ \bibinfo {pages}
  {125303} (\bibinfo {year} {2013})}\BibitemShut {NoStop}%
\bibitem [{\citenamefont {Makotyn}\ \emph {et~al.}(2014)\citenamefont
  {Makotyn}, \citenamefont {Klauss}, \citenamefont {Goldberger}, \citenamefont
  {Cornell},\ and\ \citenamefont {Jin}}]{jin14}%
  \BibitemOpen
  \bibfield  {author} {\bibinfo {author} {\bibfnamefont {P.}~\bibnamefont
  {Makotyn}}, \bibinfo {author} {\bibfnamefont {C.~E.}\ \bibnamefont {Klauss}},
  \bibinfo {author} {\bibfnamefont {D.~L.}\ \bibnamefont {Goldberger}},
  \bibinfo {author} {\bibfnamefont {E.~A.}\ \bibnamefont {Cornell}}, \ and\
  \bibinfo {author} {\bibfnamefont {D.~S.}\ \bibnamefont {Jin}},\ }\href@noop
  {} {\bibfield  {journal} {\bibinfo  {journal} {Nat. Phys.}\ }\textbf
  {\bibinfo {volume} {10}},\ \bibinfo {pages} {116} (\bibinfo {year}
  {2014})}\BibitemShut {NoStop}%
\bibitem [{\citenamefont {Sykes}\ \emph {et~al.}(2014)\citenamefont {Sykes},
  \citenamefont {Corson}, \citenamefont {D'Incao}, \citenamefont {Koller},
  \citenamefont {Greene}, \citenamefont {Rey}, \citenamefont {Hazzard},\ and\
  \citenamefont {Bohn}}]{bohn14}%
  \BibitemOpen
  \bibfield  {author} {\bibinfo {author} {\bibfnamefont {A.~G.}\ \bibnamefont
  {Sykes}}, \bibinfo {author} {\bibfnamefont {J.~P.}\ \bibnamefont {Corson}},
  \bibinfo {author} {\bibfnamefont {J.~P.}\ \bibnamefont {D'Incao}}, \bibinfo
  {author} {\bibfnamefont {A.~P.}\ \bibnamefont {Koller}}, \bibinfo {author}
  {\bibfnamefont {C.~H.}\ \bibnamefont {Greene}}, \bibinfo {author}
  {\bibfnamefont {A.~M.}\ \bibnamefont {Rey}}, \bibinfo {author} {\bibfnamefont
  {K.~R.~A.}\ \bibnamefont {Hazzard}}, \ and\ \bibinfo {author} {\bibfnamefont
  {J.~L.}\ \bibnamefont {Bohn}},\ }\href {\doibase 10.1103/PhysRevA.89.021601}
  {\bibfield  {journal} {\bibinfo  {journal} {Phys. Rev. A}\ }\textbf {\bibinfo
  {volume} {89}},\ \bibinfo {pages} {021601(R)} (\bibinfo {year}
  {2014})}\BibitemShut {NoStop}%
\bibitem [{\citenamefont {Jiang}\ \emph {et~al.}(2014)\citenamefont {Jiang},
  \citenamefont {Liu}, \citenamefont {Semenoff},\ and\ \citenamefont
  {Zhou}}]{zhou14}%
  \BibitemOpen
  \bibfield  {author} {\bibinfo {author} {\bibfnamefont {S.-J.}\ \bibnamefont
  {Jiang}}, \bibinfo {author} {\bibfnamefont {W.-M.}\ \bibnamefont {Liu}},
  \bibinfo {author} {\bibfnamefont {G.~W.}\ \bibnamefont {Semenoff}}, \ and\
  \bibinfo {author} {\bibfnamefont {F.}~\bibnamefont {Zhou}},\ }\href {\doibase
  10.1103/PhysRevA.89.033614} {\bibfield  {journal} {\bibinfo  {journal} {Phys.
  Rev. A}\ }\textbf {\bibinfo {volume} {89}},\ \bibinfo {pages} {033614}
  (\bibinfo {year} {2014})}\BibitemShut {NoStop}%
\bibitem [{\citenamefont {Smith}\ \emph {et~al.}(2014)\citenamefont {Smith},
  \citenamefont {Braaten}, \citenamefont {Kang},\ and\ \citenamefont
  {Platter}}]{platter14}%
  \BibitemOpen
  \bibfield  {author} {\bibinfo {author} {\bibfnamefont {D.~H.}\ \bibnamefont
  {Smith}}, \bibinfo {author} {\bibfnamefont {E.}~\bibnamefont {Braaten}},
  \bibinfo {author} {\bibfnamefont {D.}~\bibnamefont {Kang}}, \ and\ \bibinfo
  {author} {\bibfnamefont {L.}~\bibnamefont {Platter}},\ }\href {\doibase
  10.1103/PhysRevLett.112.110402} {\bibfield  {journal} {\bibinfo  {journal}
  {Phys. Rev. Lett.}\ }\textbf {\bibinfo {volume} {112}},\ \bibinfo {pages}
  {110402} (\bibinfo {year} {2014})}\BibitemShut {NoStop}%
\bibitem [{\citenamefont {Rossi}\ \emph {et~al.}(2014)\citenamefont {Rossi},
  \citenamefont {Salasnich}, \citenamefont {Ancilotto},\ and\ \citenamefont
  {Toigo}}]{toigo14}%
  \BibitemOpen
  \bibfield  {author} {\bibinfo {author} {\bibfnamefont {M.}~\bibnamefont
  {Rossi}}, \bibinfo {author} {\bibfnamefont {L.}~\bibnamefont {Salasnich}},
  \bibinfo {author} {\bibfnamefont {F.}~\bibnamefont {Ancilotto}}, \ and\
  \bibinfo {author} {\bibfnamefont {F.}~\bibnamefont {Toigo}},\ }\href
  {\doibase 10.1103/PhysRevA.89.041602} {\bibfield  {journal} {\bibinfo
  {journal} {Phys. Rev. A}\ }\textbf {\bibinfo {volume} {89}},\ \bibinfo
  {pages} {041602(R)} (\bibinfo {year} {2014})}\BibitemShut {NoStop}%
\bibitem [{\citenamefont {Blume}(2012)}]{blumerev12}%
  \BibitemOpen
  \bibfield  {author} {\bibinfo {author} {\bibfnamefont {D.}~\bibnamefont
  {Blume}},\ }\href {http://stacks.iop.org/0034-4885/75/i=4/a=046401}
  {\bibfield  {journal} {\bibinfo  {journal} {Rep. Prog. Phys.}\ }\textbf
  {\bibinfo {volume} {75}},\ \bibinfo {pages} {046401} (\bibinfo {year}
  {2012})}\BibitemShut {NoStop}%
\bibitem [{\citenamefont {Carlson}\ \emph {et~al.}(2003)\citenamefont
  {Carlson}, \citenamefont {Chang}, \citenamefont {Pandharipande},\ and\
  \citenamefont {Schmidt}}]{schmidt03}%
  \BibitemOpen
  \bibfield  {author} {\bibinfo {author} {\bibfnamefont {J.}~\bibnamefont
  {Carlson}}, \bibinfo {author} {\bibfnamefont {S.-Y.}\ \bibnamefont {Chang}},
  \bibinfo {author} {\bibfnamefont {V.~R.}\ \bibnamefont {Pandharipande}}, \
  and\ \bibinfo {author} {\bibfnamefont {K.~E.}\ \bibnamefont {Schmidt}},\
  }\href {\doibase 10.1103/PhysRevLett.91.050401} {\bibfield  {journal}
  {\bibinfo  {journal} {Phys. Rev. Lett.}\ }\textbf {\bibinfo {volume} {91}},\
  \bibinfo {pages} {050401} (\bibinfo {year} {2003})}\BibitemShut {NoStop}%
\bibitem [{\citenamefont {Astrakharchik}\ \emph {et~al.}(2004)\citenamefont
  {Astrakharchik}, \citenamefont {Boronat}, \citenamefont {Casulleras},\ and\
  \citenamefont {Giorgini}}]{giorgini04}%
  \BibitemOpen
  \bibfield  {author} {\bibinfo {author} {\bibfnamefont {G.~E.}\ \bibnamefont
  {Astrakharchik}}, \bibinfo {author} {\bibfnamefont {J.}~\bibnamefont
  {Boronat}}, \bibinfo {author} {\bibfnamefont {J.}~\bibnamefont {Casulleras}},
  \ and\ \bibinfo {author} {\bibfnamefont {S.}~\bibnamefont {Giorgini}},\
  }\href {\doibase 10.1103/PhysRevLett.93.200404} {\bibfield  {journal}
  {\bibinfo  {journal} {Phys. Rev. Lett.}\ }\textbf {\bibinfo {volume} {93}},\
  \bibinfo {pages} {200404} (\bibinfo {year} {2004})}\BibitemShut {NoStop}%
\bibitem [{\citenamefont {Werner}\ and\ \citenamefont
  {Castin}(2006)}]{castin06pra}%
  \BibitemOpen
  \bibfield  {author} {\bibinfo {author} {\bibfnamefont {F.}~\bibnamefont
  {Werner}}\ and\ \bibinfo {author} {\bibfnamefont {Y.}~\bibnamefont
  {Castin}},\ }\href {\doibase 10.1103/PhysRevA.74.053604} {\bibfield
  {journal} {\bibinfo  {journal} {Phys. Rev. A}\ }\textbf {\bibinfo {volume}
  {74}},\ \bibinfo {pages} {053604} (\bibinfo {year} {2006})}\BibitemShut
  {NoStop}%
\bibitem [{\citenamefont {Thomas}(1935)}]{thomas35}%
  \BibitemOpen
  \bibfield  {author} {\bibinfo {author} {\bibfnamefont {L.~H.}\ \bibnamefont
  {Thomas}},\ }\href {\doibase 10.1103/PhysRev.47.903} {\bibfield  {journal}
  {\bibinfo  {journal} {Phys. Rev.}\ }\textbf {\bibinfo {volume} {47}},\
  \bibinfo {pages} {903} (\bibinfo {year} {1935})}\BibitemShut {NoStop}%
\bibitem [{\citenamefont {Yan}\ and\ \citenamefont {Blume}(2014)}]{gaussian14}%
  \BibitemOpen
  \bibfield  {author} {\bibinfo {author} {\bibfnamefont {Y.}~\bibnamefont
  {Yan}}\ and\ \bibinfo {author} {\bibfnamefont {D.}~\bibnamefont {Blume}},\
  }\href {\doibase 10.1103/PhysRevA.90.013620} {\bibfield  {journal} {\bibinfo
  {journal} {Phys. Rev. A}\ }\textbf {\bibinfo {volume} {90}},\ \bibinfo
  {pages} {013620} (\bibinfo {year} {2014})}\BibitemShut {NoStop}%
\bibitem [{\citenamefont {Chin}(1997)}]{chin97}%
  \BibitemOpen
  \bibfield  {author} {\bibinfo {author} {\bibfnamefont {S.~A.}\ \bibnamefont
  {Chin}},\ }\href {\doibase 10.1016/S0375-9601(97)00003-0} {\bibfield
  {journal} {\bibinfo  {journal} {Phys. Lett. A}\ }\textbf {\bibinfo {volume}
  {226}},\ \bibinfo {pages} {344 } (\bibinfo {year} {1997})}\BibitemShut
  {NoStop}%
\bibitem [{\citenamefont {Jang}\ \emph {et~al.}(2001)\citenamefont {Jang},
  \citenamefont {Jang},\ and\ \citenamefont {Voth}}]{voth}%
  \BibitemOpen
  \bibfield  {author} {\bibinfo {author} {\bibfnamefont {S.}~\bibnamefont
  {Jang}}, \bibinfo {author} {\bibfnamefont {S.}~\bibnamefont {Jang}}, \ and\
  \bibinfo {author} {\bibfnamefont {G.~A.}\ \bibnamefont {Voth}},\ }\href
  {\doibase 10.1063/1.1410117} {\bibfield  {journal} {\bibinfo  {journal} {J.
  Chem. Phys.}\ }\textbf {\bibinfo {volume} {115}},\ \bibinfo {pages} {7832}
  (\bibinfo {year} {2001})}\BibitemShut {NoStop}%
\bibitem [{Note3()}]{Note3}%
  \BibitemOpen
  \bibinfo {note} {The ground state wave function is obtained by transforming
  to hyperspherical coordinates. The hyperangular wave function, which
  separates from the hyperradial portion, is known analytically~\cite
  {Efimov,BraatenHammerReview} and the hyperradial wave function is found
  numerically.}\BibitemShut {Stop}%
\bibitem [{\citenamefont {von Stecher}(2010)}]{stecher10}%
  \BibitemOpen
  \bibfield  {author} {\bibinfo {author} {\bibfnamefont {J.}~\bibnamefont {von
  Stecher}},\ }\href {http://stacks.iop.org/0953-4075/43/i=10/a=101002}
  {\bibfield  {journal} {\bibinfo  {journal} {J. Phys. B}\ }\textbf {\bibinfo
  {volume} {43}},\ \bibinfo {pages} {101002} (\bibinfo {year}
  {2010})}\BibitemShut {NoStop}%
\bibitem [{\citenamefont {Nicholson}(2012)}]{amy12}%
  \BibitemOpen
  \bibfield  {author} {\bibinfo {author} {\bibfnamefont {A.~N.}\ \bibnamefont
  {Nicholson}},\ }\href {\doibase 10.1103/PhysRevLett.109.073003} {\bibfield
  {journal} {\bibinfo  {journal} {Phys. Rev. Lett.}\ }\textbf {\bibinfo
  {volume} {109}},\ \bibinfo {pages} {073003} (\bibinfo {year}
  {2012})}\BibitemShut {NoStop}%
\bibitem [{\citenamefont {Gattobigio}\ and\ \citenamefont
  {Kievsky}(2014)}]{kievsky14}%
  \BibitemOpen
  \bibfield  {author} {\bibinfo {author} {\bibfnamefont {M.}~\bibnamefont
  {Gattobigio}}\ and\ \bibinfo {author} {\bibfnamefont {A.}~\bibnamefont
  {Kievsky}},\ }\href {\doibase 10.1103/PhysRevA.90.012502} {\bibfield
  {journal} {\bibinfo  {journal} {Phys. Rev. A}\ }\textbf {\bibinfo {volume}
  {90}},\ \bibinfo {pages} {012502} (\bibinfo {year} {2014})}\BibitemShut
  {NoStop}%
\bibitem [{Note4()}]{Note4}%
  \BibitemOpen
  \bibinfo {note} {Y. Yan and D. Blume, in preparation.}\BibitemShut {Stop}%
\bibitem [{\citenamefont {Kogut}(1983)}]{kogut83}%
  \BibitemOpen
  \bibfield  {author} {\bibinfo {author} {\bibfnamefont {J.~B.}\ \bibnamefont
  {Kogut}},\ }\href {\doibase 10.1103/RevModPhys.55.775} {\bibfield  {journal}
  {\bibinfo  {journal} {Rev. Mod. Phys.}\ }\textbf {\bibinfo {volume} {55}},\
  \bibinfo {pages} {775} (\bibinfo {year} {1983})}\BibitemShut {NoStop}%
\bibitem [{\citenamefont {Lee}(2009)}]{lee09}%
  \BibitemOpen
  \bibfield  {author} {\bibinfo {author} {\bibfnamefont {D.}~\bibnamefont
  {Lee}},\ }\href {\doibase 10.1016/j.ppnp.2008.12.001} {\bibfield  {journal}
  {\bibinfo  {journal} {Progress in Particle and Nuclear Physics}\ }\textbf
  {\bibinfo {volume} {63}},\ \bibinfo {pages} {117 } (\bibinfo {year}
  {2009})}\BibitemShut {NoStop}%
\bibitem [{\citenamefont {Drut}\ and\ \citenamefont {Nicholson}(2013)}]{amy13}%
  \BibitemOpen
  \bibfield  {author} {\bibinfo {author} {\bibfnamefont {J.~E.}\ \bibnamefont
  {Drut}}\ and\ \bibinfo {author} {\bibfnamefont {A.~N.}\ \bibnamefont
  {Nicholson}},\ }\href {\doibase 10.1088/0954-3899/40/4/043101} {\bibfield
  {journal} {\bibinfo  {journal} {J. Phys. G}\ }\textbf {\bibinfo {volume}
  {40}},\ \bibinfo {pages} {043101} (\bibinfo {year} {2013})}\BibitemShut
  {NoStop}%
\bibitem [{\citenamefont {Esry}\ and\ \citenamefont {Greene}(1999)}]{esry99}%
  \BibitemOpen
  \bibfield  {author} {\bibinfo {author} {\bibfnamefont {B.~D.}\ \bibnamefont
  {Esry}}\ and\ \bibinfo {author} {\bibfnamefont {C.~H.}\ \bibnamefont
  {Greene}},\ }\href {\doibase 10.1103/PhysRevA.60.1451} {\bibfield  {journal}
  {\bibinfo  {journal} {Phys. Rev. A}\ }\textbf {\bibinfo {volume} {60}},\
  \bibinfo {pages} {1451} (\bibinfo {year} {1999})}\BibitemShut {NoStop}%
\bibitem [{\citenamefont {Alhassid}\ \emph {et~al.}(2008)\citenamefont
  {Alhassid}, \citenamefont {Bertsch},\ and\ \citenamefont
  {Fang}}]{alhassid08}%
  \BibitemOpen
  \bibfield  {author} {\bibinfo {author} {\bibfnamefont {Y.}~\bibnamefont
  {Alhassid}}, \bibinfo {author} {\bibfnamefont {G.~F.}\ \bibnamefont
  {Bertsch}}, \ and\ \bibinfo {author} {\bibfnamefont {L.}~\bibnamefont
  {Fang}},\ }\href {\doibase 10.1103/PhysRevLett.100.230401} {\bibfield
  {journal} {\bibinfo  {journal} {Phys. Rev. Lett.}\ }\textbf {\bibinfo
  {volume} {100}},\ \bibinfo {pages} {230401} (\bibinfo {year}
  {2008})}\BibitemShut {NoStop}%
\bibitem [{\citenamefont {Rontani}\ \emph {et~al.}(2009)\citenamefont
  {Rontani}, \citenamefont {Armstrong}, \citenamefont {Yu}, \citenamefont
  {\AA{}berg},\ and\ \citenamefont {Reimann}}]{reimann09}%
  \BibitemOpen
  \bibfield  {author} {\bibinfo {author} {\bibfnamefont {M.}~\bibnamefont
  {Rontani}}, \bibinfo {author} {\bibfnamefont {J.~R.}\ \bibnamefont
  {Armstrong}}, \bibinfo {author} {\bibfnamefont {Y.}~\bibnamefont {Yu}},
  \bibinfo {author} {\bibfnamefont {S.}~\bibnamefont {\AA{}berg}}, \ and\
  \bibinfo {author} {\bibfnamefont {S.~M.}\ \bibnamefont {Reimann}},\ }\href
  {\doibase 10.1103/PhysRevLett.102.060401} {\bibfield  {journal} {\bibinfo
  {journal} {Phys. Rev. Lett.}\ }\textbf {\bibinfo {volume} {102}},\ \bibinfo
  {pages} {060401} (\bibinfo {year} {2009})}\BibitemShut {NoStop}%
\bibitem [{\citenamefont {Werner}\ and\ \citenamefont {Castin}(2012)}]{Yvan12}%
  \BibitemOpen
  \bibfield  {author} {\bibinfo {author} {\bibfnamefont {F.}~\bibnamefont
  {Werner}}\ and\ \bibinfo {author} {\bibfnamefont {Y.}~\bibnamefont
  {Castin}},\ }\href {\doibase 10.1103/PhysRevA.86.013626} {\bibfield
  {journal} {\bibinfo  {journal} {Phys. Rev. A}\ }\textbf {\bibinfo {volume}
  {86}},\ \bibinfo {pages} {013626} (\bibinfo {year} {2012})}\BibitemShut
  {NoStop}%
\bibitem [{\citenamefont {Braaten}\ and\ \citenamefont
  {Platter}(2008)}]{platter08}%
  \BibitemOpen
  \bibfield  {author} {\bibinfo {author} {\bibfnamefont {E.}~\bibnamefont
  {Braaten}}\ and\ \bibinfo {author} {\bibfnamefont {L.}~\bibnamefont
  {Platter}},\ }\href {\doibase 10.1103/PhysRevLett.100.205301} {\bibfield
  {journal} {\bibinfo  {journal} {Phys. Rev. Lett.}\ }\textbf {\bibinfo
  {volume} {100}},\ \bibinfo {pages} {205301} (\bibinfo {year}
  {2008})}\BibitemShut {NoStop}%
\bibitem [{\citenamefont {Yu}\ \emph {et~al.}(2009)\citenamefont {Yu},
  \citenamefont {Bruun},\ and\ \citenamefont {Baym}}]{gordon09}%
  \BibitemOpen
  \bibfield  {author} {\bibinfo {author} {\bibfnamefont {Z.}~\bibnamefont
  {Yu}}, \bibinfo {author} {\bibfnamefont {G.~M.}\ \bibnamefont {Bruun}}, \
  and\ \bibinfo {author} {\bibfnamefont {G.}~\bibnamefont {Baym}},\ }\href
  {\doibase 10.1103/PhysRevA.80.023615} {\bibfield  {journal} {\bibinfo
  {journal} {Phys. Rev. A}\ }\textbf {\bibinfo {volume} {80}},\ \bibinfo
  {pages} {023615} (\bibinfo {year} {2009})}\BibitemShut {NoStop}%
\bibitem [{\citenamefont {Drut}\ \emph {et~al.}(2011)\citenamefont {Drut},
  \citenamefont {L\"ahde},\ and\ \citenamefont {Ten}}]{timour11}%
  \BibitemOpen
  \bibfield  {author} {\bibinfo {author} {\bibfnamefont {J.~E.}\ \bibnamefont
  {Drut}}, \bibinfo {author} {\bibfnamefont {T.~A.}\ \bibnamefont {L\"ahde}}, \
  and\ \bibinfo {author} {\bibfnamefont {T.}~\bibnamefont {Ten}},\ }\href
  {\doibase 10.1103/PhysRevLett.106.205302} {\bibfield  {journal} {\bibinfo
  {journal} {Phys. Rev. Lett.}\ }\textbf {\bibinfo {volume} {106}},\ \bibinfo
  {pages} {205302} (\bibinfo {year} {2011})}\BibitemShut {NoStop}%
\bibitem [{\citenamefont {Doggen}\ and\ \citenamefont
  {Kinnunen}(2013)}]{doggen13}%
  \BibitemOpen
  \bibfield  {author} {\bibinfo {author} {\bibfnamefont {E.~V.~H.}\
  \bibnamefont {Doggen}}\ and\ \bibinfo {author} {\bibfnamefont {J.~J.}\
  \bibnamefont {Kinnunen}},\ }\href {\doibase 10.1103/PhysRevLett.111.025302}
  {\bibfield  {journal} {\bibinfo  {journal} {Phys. Rev. Lett.}\ }\textbf
  {\bibinfo {volume} {111}},\ \bibinfo {pages} {025302} (\bibinfo {year}
  {2013})}\BibitemShut {NoStop}%
\bibitem [{\citenamefont {Yan}\ and\ \citenamefont {Blume}(2013)}]{contact13}%
  \BibitemOpen
  \bibfield  {author} {\bibinfo {author} {\bibfnamefont {Y.}~\bibnamefont
  {Yan}}\ and\ \bibinfo {author} {\bibfnamefont {D.}~\bibnamefont {Blume}},\
  }\href {\doibase 10.1103/PhysRevA.88.023616} {\bibfield  {journal} {\bibinfo
  {journal} {Phys. Rev. A}\ }\textbf {\bibinfo {volume} {88}},\ \bibinfo
  {pages} {023616} (\bibinfo {year} {2013})}\BibitemShut {NoStop}%
\bibitem [{\citenamefont {Stewart}\ \emph {et~al.}(2010)\citenamefont
  {Stewart}, \citenamefont {Gaebler}, \citenamefont {Drake},\ and\
  \citenamefont {Jin}}]{jin10}%
  \BibitemOpen
  \bibfield  {author} {\bibinfo {author} {\bibfnamefont {J.~T.}\ \bibnamefont
  {Stewart}}, \bibinfo {author} {\bibfnamefont {J.~P.}\ \bibnamefont
  {Gaebler}}, \bibinfo {author} {\bibfnamefont {T.~E.}\ \bibnamefont {Drake}},
  \ and\ \bibinfo {author} {\bibfnamefont {D.~S.}\ \bibnamefont {Jin}},\ }\href
  {\doibase 10.1103/PhysRevLett.104.235301} {\bibfield  {journal} {\bibinfo
  {journal} {Phys. Rev. Lett.}\ }\textbf {\bibinfo {volume} {104}},\ \bibinfo
  {pages} {235301} (\bibinfo {year} {2010})}\BibitemShut {NoStop}%
\bibitem [{\citenamefont {Kuhnle}\ \emph {et~al.}(2010)\citenamefont {Kuhnle},
  \citenamefont {Hu}, \citenamefont {Liu}, \citenamefont {Dyke}, \citenamefont
  {Mark}, \citenamefont {Drummond}, \citenamefont {Hannaford},\ and\
  \citenamefont {Vale}}]{vale10}%
  \BibitemOpen
  \bibfield  {author} {\bibinfo {author} {\bibfnamefont {E.~D.}\ \bibnamefont
  {Kuhnle}}, \bibinfo {author} {\bibfnamefont {H.}~\bibnamefont {Hu}}, \bibinfo
  {author} {\bibfnamefont {X.-J.}\ \bibnamefont {Liu}}, \bibinfo {author}
  {\bibfnamefont {P.}~\bibnamefont {Dyke}}, \bibinfo {author} {\bibfnamefont
  {M.}~\bibnamefont {Mark}}, \bibinfo {author} {\bibfnamefont {P.~D.}\
  \bibnamefont {Drummond}}, \bibinfo {author} {\bibfnamefont {P.}~\bibnamefont
  {Hannaford}}, \ and\ \bibinfo {author} {\bibfnamefont {C.~J.}\ \bibnamefont
  {Vale}},\ }\href {\doibase 10.1103/PhysRevLett.105.070402} {\bibfield
  {journal} {\bibinfo  {journal} {Phys. Rev. Lett.}\ }\textbf {\bibinfo
  {volume} {105}},\ \bibinfo {pages} {070402} (\bibinfo {year}
  {2010})}\BibitemShut {NoStop}%
\bibitem [{\citenamefont {Kuhnle}\ \emph {et~al.}(2011)\citenamefont {Kuhnle},
  \citenamefont {Hoinka}, \citenamefont {Dyke}, \citenamefont {Hu},
  \citenamefont {Hannaford},\ and\ \citenamefont {Vale}}]{vale11}%
  \BibitemOpen
  \bibfield  {author} {\bibinfo {author} {\bibfnamefont {E.~D.}\ \bibnamefont
  {Kuhnle}}, \bibinfo {author} {\bibfnamefont {S.}~\bibnamefont {Hoinka}},
  \bibinfo {author} {\bibfnamefont {P.}~\bibnamefont {Dyke}}, \bibinfo {author}
  {\bibfnamefont {H.}~\bibnamefont {Hu}}, \bibinfo {author} {\bibfnamefont
  {P.}~\bibnamefont {Hannaford}}, \ and\ \bibinfo {author} {\bibfnamefont
  {C.~J.}\ \bibnamefont {Vale}},\ }\href {\doibase
  10.1103/PhysRevLett.106.170402} {\bibfield  {journal} {\bibinfo  {journal}
  {Phys. Rev. Lett.}\ }\textbf {\bibinfo {volume} {106}},\ \bibinfo {pages}
  {170402} (\bibinfo {year} {2011})}\BibitemShut {NoStop}%
\bibitem [{\citenamefont {Sagi}\ \emph {et~al.}(2012)\citenamefont {Sagi},
  \citenamefont {Drake}, \citenamefont {Paudel},\ and\ \citenamefont
  {Jin}}]{jin12}%
  \BibitemOpen
  \bibfield  {author} {\bibinfo {author} {\bibfnamefont {Y.}~\bibnamefont
  {Sagi}}, \bibinfo {author} {\bibfnamefont {T.~E.}\ \bibnamefont {Drake}},
  \bibinfo {author} {\bibfnamefont {R.}~\bibnamefont {Paudel}}, \ and\ \bibinfo
  {author} {\bibfnamefont {D.~S.}\ \bibnamefont {Jin}},\ }\href {\doibase
  10.1103/PhysRevLett.109.220402} {\bibfield  {journal} {\bibinfo  {journal}
  {Phys. Rev. Lett.}\ }\textbf {\bibinfo {volume} {109}},\ \bibinfo {pages}
  {220402} (\bibinfo {year} {2012})}\BibitemShut {NoStop}%
\bibitem [{\citenamefont {Blume}\ and\ \citenamefont {Daily}(2009)}]{daily09}%
  \BibitemOpen
  \bibfield  {author} {\bibinfo {author} {\bibfnamefont {D.}~\bibnamefont
  {Blume}}\ and\ \bibinfo {author} {\bibfnamefont {K.~M.}\ \bibnamefont
  {Daily}},\ }\href {\doibase 10.1103/PhysRevA.80.053626} {\bibfield  {journal}
  {\bibinfo  {journal} {Phys. Rev. A}\ }\textbf {\bibinfo {volume} {80}},\
  \bibinfo {pages} {053626} (\bibinfo {year} {2009})}\BibitemShut {NoStop}%
\end{thebibliography}
  \end{document}